\documentclass[aps,pre,twocolumn,groupedaddress,amsmath,amssymb]{revtex4-1}
\usepackage{tikz}
\usetikzlibrary{shapes}
\usepackage{graphicx}  
\usepackage{dcolumn}   
\usepackage{bm}        
\usepackage{verbatim}   
\usepackage{mathrsfs}  
\graphicspath{{./figures/}}
\newcolumntype{M}[1]{>{\centering\arraybackslash}m{#1}}
\newcolumntype{N}{@{}m{0pt}@{}}
\usepackage{multirow}
\newcommand{\uFluct}{u'}
\newcommand{\pFluct}{p'}
\newcommand{\rhoFluct}{\rho '}
\newcommand{\sFluct}{s'}
\newcommand{\TFluct}{T'}
\newcommand{\revision}[1]{{\color{black}#1}}
\usepackage[T1]{fontenc}

\newcommand{\markerone}{\raisebox{0.5pt}{\tikz{\node[draw,scale=0.4,circle,fill=black!20!black](){};}}}
\newcommand{\markertwo}{\raisebox{0.5pt}{\tikz{\node[draw,scale=0.4,circle,fill=black!20!gray](){};}}}
\newcommand{\markerthree}{\raisebox{0.5pt}{\tikz{\node[draw,scale=0.4,circle,fill=none](){};}}}
\newcommand{\markerfour}{\raisebox{0.5pt}{\tikz{\node[draw,scale=0.4,regular polygon, regular polygon sides=4,fill=black!20!black](){};}}}
\newcommand{\markerfive}{\raisebox{0.5pt}{\tikz{\node[draw,scale=0.4,regular polygon, regular polygon sides=4,fill=black!20!gray](){};}}}
\newcommand{\markersix}{\raisebox{0.5pt}{\tikz{\node[draw,scale=0.4,regular polygon, regular polygon sides=4,fill=none](){};}}}
\usepackage{footmisc}

\usepackage{scalerel,stackengine}
\stackMath
\newcommand\reallywidehat[1]{%
\savestack{\tmpbox}{\stretchto{%
  \scaleto{%
    \scalerel*[\widthof{\ensuremath{#1}}]{\kern-.6pt\bigwedge\kern-.6pt}%
    {\rule[-\textheight/2]{1ex}{\textheight}}
  }{\textheight}%
}{0.5ex}}%
\stackon[1pt]{#1}{\tmpbox}%
}
\begin{document}

\title{Spectral energy cascade and decay in nonlinear acoustic waves}
\author{Prateek Gupta}
\email[]{gupta288@purdue.edu}
\affiliation{
  School of Mechanical Engineering,\\
   Purdue University,\\
  West Lafayette, IN 47906, USA
   }
\author{Carlo Scalo}
\affiliation{
  School of Mechanical Engineering,\\
   Purdue University,\\
  West Lafayette, IN 47906, USA
   }
   
\date{\today}
\begin{abstract}
{
We present a numerical and theoretical investigation of nonlinear spectral energy cascade of decaying finite-amplitude planar acoustic waves in a single-component ideal gas at standard temperature and pressure (STP). We analyze various one-dimensional canonical flow configurations: a propagating traveling wave (TW), a standing wave (SW), and randomly initialized Acoustic Wave Turbulence (AWT). Due to nonlinear wave propagation, energy at the large scales cascades down to smaller scales dominated by viscous dissipation, analogous to hydrodynamic turbulence. We use shock-resolved mesh-adaptive direct numerical simulations (DNS) of the fully compressible one-dimensional Navier-Stokes equations to simulate the spectral energy cascade in nonlinear acoustic waves. The simulation parameter space for the TW, SW, and AWT cases spans three orders of magnitude in initial wave pressure amplitude and dynamic viscosity, thus covering a wide range of both spectral energy cascade and the viscous dissipation rates. \revision{The shock waves formed as a result of energy cascade are weak ($M$ $<1.4$), and hence we neglect thermodynamic non-equilibrium effects such as molecular vibrational relaxation in the current study.} We also derive a new set of nonlinear acoustics equations truncated to second order and the corresponding perturbation energy corollary yielding the expression for a new perturbation energy norm $E^{(2)}$. Its spatial average, <$E^{(2)}$> satisfies the definition of a Lyapunov function, correctly capturing the inviscid (or lossless) broadening of spectral energy in the initial stages of evolution -- analogous to the evolution of kinetic energy during the hydrodynamic break down of three-dimensional coherent vorticity -- resulting in the formation of smaller scales. Upon saturation of the spectral energy cascade i.e. fully broadened energy spectrum, the onset of viscous losses causes a monotonic decay of <$E^{(2)}$> in time. In this regime, the DNS results yield <$E^{(2)}$> $\sim t^{-2}$ for TW and SW, and <$E^{(2)}$> $\sim t^{-2/3}$ for AWT initialized with white noise. Using the perturbation energy corollary, we derive analytical expressions for the energy, energy flux, and dissipation rate in the wavenumber space. These yield the definitions of characteristic length scales such as the integral length scale $\ell$ (characteristic initial energy containing scale) and the Kolmogorov length scale $\eta$ (shock thickness scale), analogous to K41 theory of hydrodynamic turbulence (A. N. Kolmogorov, \emph{Dokl. Akad. Nauk SSSR} 30 , 9 (1941)). Finally, we show that the fully developed energy spectrum of the nonlinear acoustic waves scales as $\widehat{E}_k k^2\epsilon^{-2/3}\ell^{1/3}\sim C\,f(k\eta)$, with $C \approx 0.075$ constant for TW and SW but decaying in time for AWT.
}
 \end{abstract}
\maketitle

\section{Introduction}

{Nonlinear wave processes are observed in a variety of engineering and physics applications such as acoustics~\citep{Hamilton_NLA_1998, Lighthill_JSV_1978}, combustion noise~\citep{Bonciolini_PRE_2017, DouasbinEtAl_JCP_2018}, jet noise~\citep{Baars_JETNOISE_2014, Von_PRE_2012, Baars_ExpFluids_2013}, thermoacoustics~\citep{ScaloLH_JFM_2015, GuptaLS_JFM_2017}, surface waves~\citep{AblowitzClarkson1991}, and plasma-physics~\citep{Gurbatov2012}, requiring nonlinear evolution equations to describe the dynamics of perturbations}. In the case of high amplitude planar acoustic wave propagation, two main nonlinear effects are present: acoustic streaming~\citep{Lighthill_JSV_1978, Gedeon_1997_Cryocoolers} and wave steepening~\citep{Hamilton_NLA_1998, Naugolnykh1998}. Acoustic streaming is an Eulerian mean flow and is attributed to the kinematic nonlinearities~\cite{Lighthill_JSV_1978}. Convective derivatives of velocity in momentum conservation equation cause wave induced Reynolds stresses~\citep{ScaloLH_JFM_2015} which have non-zero mean values in time. In one dimension, longitudinal stresses are generated which cause steady mass flow due to wave propagation. On the other hand, wave steepening occurs due to local gradients in the wave speed associated with thermodynamic nonlinearities~\citep{Hamilton_NLA_1998}. Wave steepening entails generation of smaller length scales via a nonlinear energy cascade, {which can be exemplified by developing the product of two truncated Fourier series,}
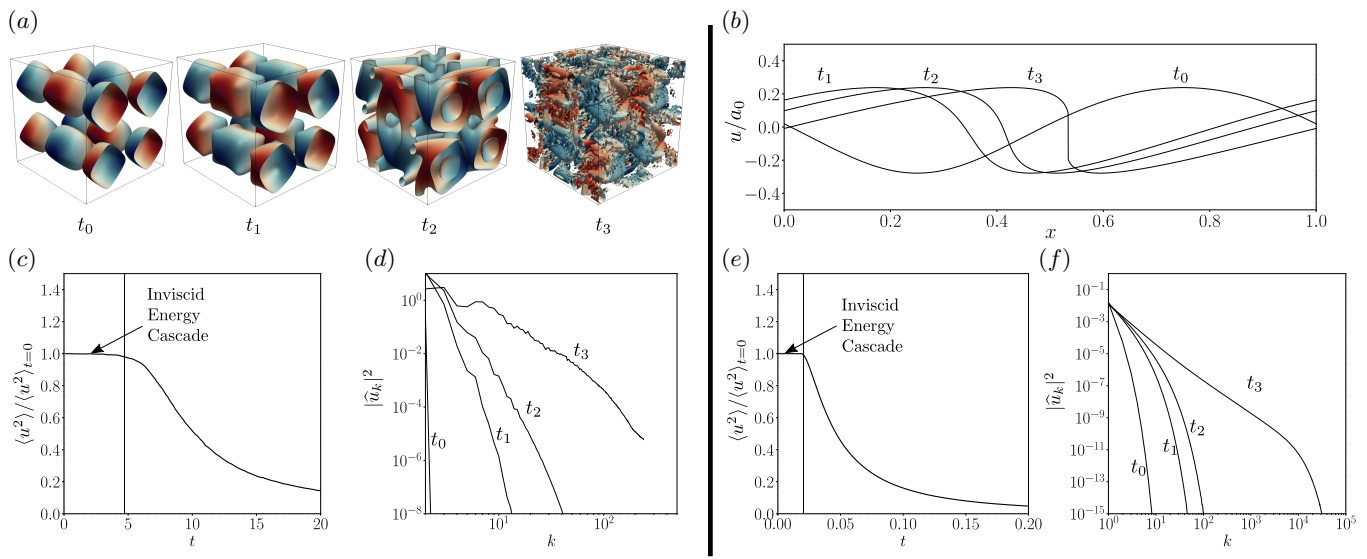
\begin{figure*}[!t]
\centering
\includegraphics[width=\linewidth]{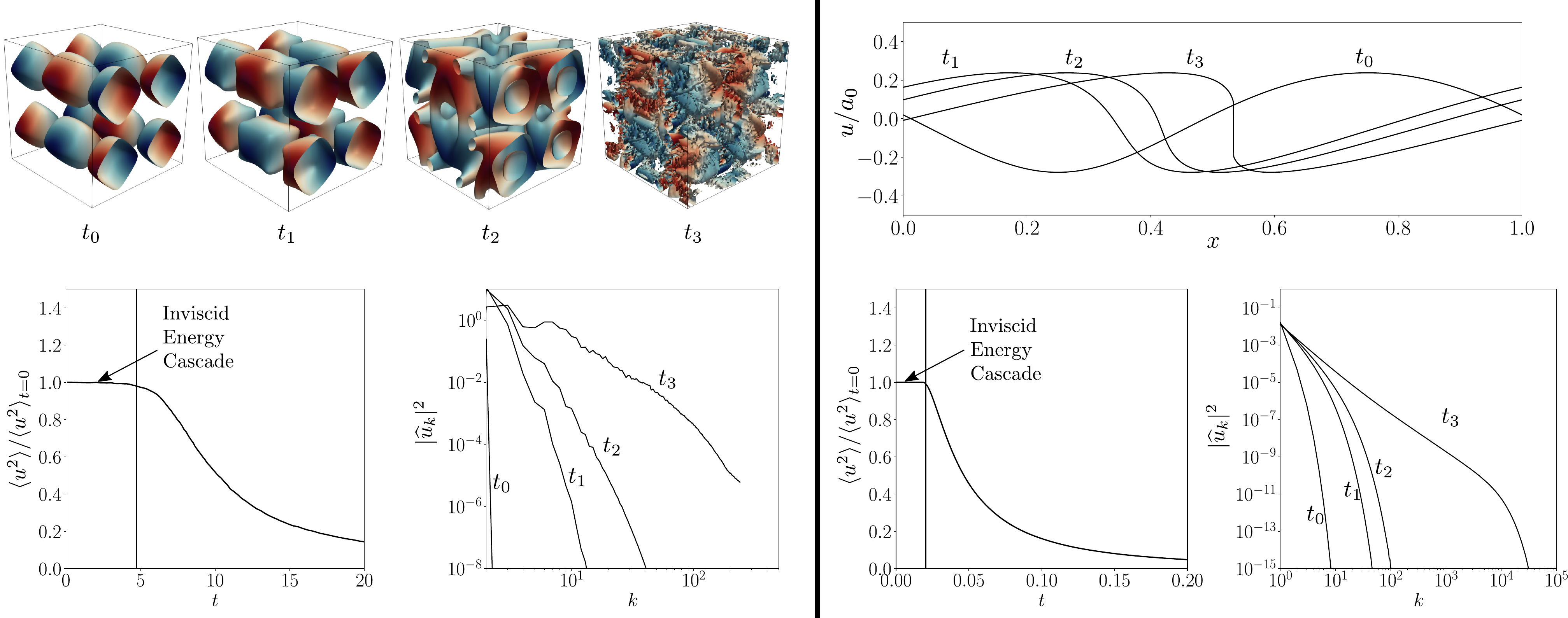}
\put(-510,200){$(a)$}
\put(-240,200){$(b)$}
\put(-510,110){$(c)$}
\put(-375,110){$(d)$}
\put(-240,110){$(e)$}
\put(-120,110){$(f)$}
\caption{ $Q$-criterion iso-surfaces colored with the local velocity magnitude obtained from a direct numerical simulation of a Taylor-Green vortex in a triply-periodic domain $\left[-\pi, \pi\right]^3$~\cite{Chapelier_jcp_2017}, exhibiting breakdown into hydrodynamic turbulence $(a)$, velocity perturbation field in a high amplitude nonlinear traveling acoustic wave (TW) $(b)$, evolution of normalized spatial average of $u^2$ $(c),(e)$, and velocity spectra $|\widehat{u}_k|^2$ $(d), (f)$ at times $t_0, t_1, t_2, t_3$. The spectral broadening occurs due to the nonlinear terms in the governing equations generating smaller length scales resulting in energy cascade from larger to smaller length scales.
}
\label{fig: SpectraTurbulenceNLA}
\end{figure*}
\begin{eqnarray}
\left(\sum^{n}_{k=-n}a_k e^{2\pi ikx}\right)\left(\sum^{m}_{l=-m}b_le^{2\pi ilx}\right) = \nonumber\\
={\sum_k a_k b_{-k}} +{\underset{k+l\neq 0}{\sum_k \sum_l} a_k b_l e^{2\pi i(k+l)x}}.
\label{eq: IntroTrig}
\end{eqnarray}
The left hand side of the Eq.~\eqref{eq: IntroTrig} represents a generic quadratic nonlinear term appearing in a governing equation. Continued nonlinear evolution results in further generation of smaller length scales, as depicted by the second term on the right hand side of Eq.~\eqref{eq: IntroTrig}, ultimately leading to spectral broadening. In the case of nonlinear acoustic waves, the shock thickness is the smallest length scale present in the flow, governed by the viscous dissipation. {The latter causes saturation of the spectral broadening process, hence establishing an energy flow (primarily) directed from large scales to small scales. Identical spectral energy dynamics are observed in classic hydrodynamic turbulence~\cite{tennekes1972first}, where nonlinear processes such as vortex stretching and tilting (only existing in three-dimensions) cause spontaneous generation of progressively smaller vortical structures (i.e. eddies), until velocity gradients become sufficiently large for viscous dissipation to become relevant (see Figure~\ref{fig: SpectraTurbulenceNLA}).}

{In previous numerical investigations~\citep{GuptaLS_JFM_2017} -- inspired by the experimental setups in \cite{BiwaEtAl_JASA_2014, YazakiIMT_PhysRevLett_1998} -- the present authors have demonstrated the existence of an equilibrium spectral energy cascade in quasi-planar weak shock waves sustained by thermoacoustic instabilities in a resonator. The latter inject energy only at scales comparable to the resonator length (large scales); harmonic generation then takes place, leading to spectral broadening and progressive generation of smaller scales until viscous losses, occurring at the shock-thickness scale, dominate the energy cascade. Building upon the findings in \citep{GuptaLS_JFM_2017}, in the present work, we mathematically formalize the dynamics of nonlinear acoustic spectral energy cascade in a more canonical setup neglecting thermoacoustic energy sources and focusing on purely planar waves.

Due to the absence of physical sources of energy, the energy of nonlinear acoustic waves (if correctly defined) decays monotonically in time due to viscous dissipation at small length scales, analogous to freely decaying hydrodynamic turbulence (see Figure~\ref{fig: SpectraTurbulenceNLA})}. We study the spatio-temporal and spectro-temporal evolution of such finite amplitude planar nonlinear acoustic waves in three canonical configurations in particular: traveling waves (TW), standing waves (SW), and randomly initialized Acoustic Wave Turbulence (AWT). {In spite of it's theoretical nature, planar nonlinear acoustic theory is still commonly used in practical investigations such as sonic boom propagation (TW)~\cite{Crow_JFM_1969}, rotating detonation engines (SW)~\cite{Schwinn_CandF_2018}, combustion chamber noise (AWT)~\cite{Culik_AGARD_2006}, and thermoacoustics ~\cite{BiwaEtAl_JASA_2014, YazakiIMT_PhysRevLett_1998}.} We study these configurations for pressure amplitudes and viscosities spanning three orders of magnitudes. Utilizing the second order nonlinear acoustics approximation, we derive analytical expressions for the spectral energy, energy transfer function, and dissipation. Analogous to the study of small scale generation in hydrodynamic turbulence, well quantified by the K41 theory~\citep{kolmogorov1941a, kolmogorov1941b, kolmogorov1941c}, we also define the relevant length scales associated to fully developed nonlinear acoustic waves elucidating the scaling features of the energy spectra. To this end, we perform the direct numerical simulations (DNS) resolving all the relevant length scales~\cite{pope2000turbulent} of nonlinear acoustic wave propagation.

Usually, problems in nondispersive nonlinear wave propagation are studied utilizing the model Burgers equation \citep{Whitham2011, Gurbatov_1981_JETP, Naugolnykh1998, Gurbatov2012}. The spectral energy and decay dynamics of one dimensional Burgers turbulence have been studied extensively by Kida~\citep{Kida_1979_JFM}, Gurbatov et al.~\citep{Gurbatov_JFM_1997, Gurbatov1991}, Woyczynski~\citep{Woyczynski_Gottingen_2006}, Fournier and Frisch~\cite{Fournier1983Burgers}, and Burgers~\cite{Burgers1974nonlinear}. 
However, the equations of second order nonlinear acoustics can be reduced to Burgers equation only assuming planar TW, thus limiting its applicability. Generalized problems involving an ensemble of acoustic waves of different amplitudes, such as AWT, have also been subjects of detailed analysis~\citep{Kida_1979_JFM, Newell_2011_ARFM}. Such studies primarily involve formulation of the kinetic equations of complex amplitudes of weakly nonlinear harmonic waves~\cite{Zakharov2012}. Utilizing the kinetic equations wave interaction potentials are defined in the context of wave-wave interactions. However, such analysis are restricted to complex harmonic representation of waves in space and time and hence, fail to elucidate the inter-scale energy transfer dynamics due to general nonlinear wave interactions.{\color{black}{ In this work, we utilize the continuum gas dynamics governing equations to elucidate the spectral energy cascade and decay dynamics of nonlinear acoustics. The nonlinear equations governing high amplitude acoustics yield novel analytical expressions for spectral energy, spectral energy flux, and spectral dissipation rate valid for planar nonlinear acoustic waves with general phasing}}. The dissipation causes power law decay of energy in time due to gradual increase of the dissipative length scale. Such decay dynamics occur due to separation of energy containing and diffusive length scales and resemble those of decaying homogeneous isotropic turbulence (HIT) \citep{batchelor1948decay, batchelor1949nature, hinze1975turbulence, Ishihara_ARFM_2009}.

We present a framework for studying nonlinear acoustic wave propagation phenomenon in one dimension utilizing the second order nonlinear acoustics equations and DNS of compressible 1D Navier-Stokes (resolving all length scales). We derive the former utilizing the entropy scaling considerations for weak shocks as discussed in Section~\ref{sec: EntropyScaling}. In Section~\ref{sec: EnergyCorollary}, we derive a novel perturbation energy corollary for nonlinear acoustic perturbations utilizing the second order governing equations yielding a new perturbation energy function. Its spatial average defines the Lyapunov function of the system and decays monotonically in the presence of dissipation and absence of energy sources which is confirmed through the DNS data shown in Section~\ref{sec: numerics} along with brief explaination of numerical technique utilized. In Section~\ref{sec: ScalesAcousticCascade}, we derive the spectral energy conservation equation thus identifying the spectral energy flux and spectral dissipation utilizing the energy corollary. Furthermore, we discuss the evolution of the primary length scales involved in the spectral energy cascade and decay. Finally, in Section~\ref{sec: SpectralScaling}, we show the scaling of spectral energy, spectral energy flux, and spectral dissipation. Throughout, the theoretical results are supported utilizing the DNS of three specific cases of acoustic waves namely, single harmonic traveling wave (TW), single harmonic standing wave (SW), and random broadband noise (AWT). 
\section{Governing equations and scaling analysis}
\label{sec: EntropyScaling}
{In this section, we derive the governing equations for nonlinear acoustics truncated up to second order (in the acoustic perturbation variables) for a single-component ideal gas. We begin with fully compressible one-dimensional Navier-Stokes equations for continuum gas dynamics and analysis of entropy scaling with pressure jumps in weak shocks formed due to the steepening of nonlinear acoustic waves~(Section~\ref{sec: 1DNS}). We then briefly discuss the variable decomposition and non-dimensionalization in 
Section~\ref{sec: 2ndOrderEntropy}, followed by the derivation of second order governing equations for nonlinear acoustics in Section~\ref{sec: 2ndOrderNLA}.}

\subsection{Fully compressible 1D Navier-Stokes and entropy scaling in weak shocks}
\label{sec: 1DNS}
\begin{figure}[!b]
\centering
\includegraphics[width=\linewidth]{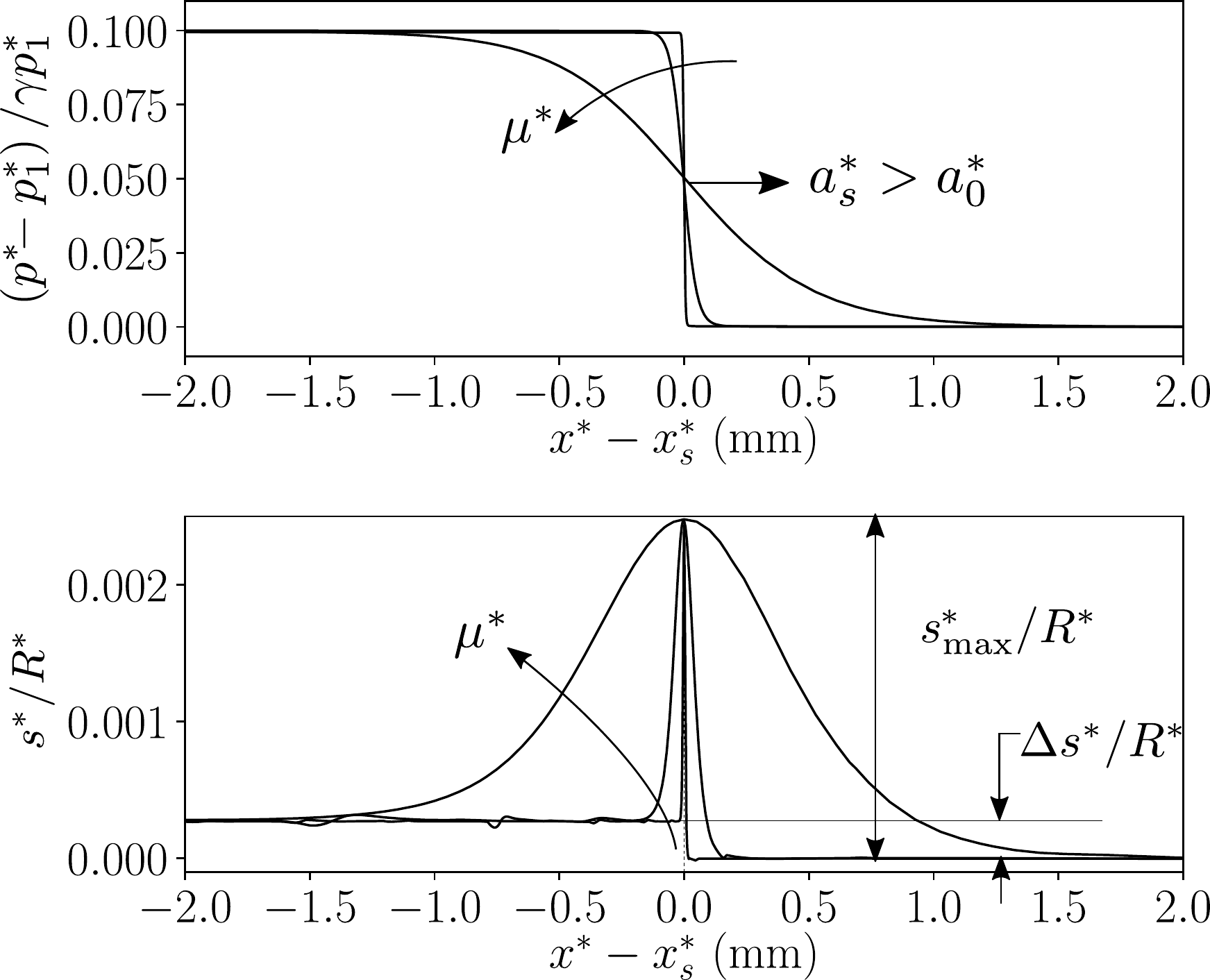}
\put(-250,200){$(a)$}
\put(-250,97){$(b)$}
\caption{Weak shock wave structure $(a)$ pressure $p^*$ and $(b)$ entropy $s^*$ propagating with a speed $a^*_s>a^*_0$ obtained from DNS (see Section~\ref{sec: numerics}). $\Delta s^*/R^*$ and $s^*_{\mathrm{max}}/R^*$ are the entropy jump and maximum entropy respectively. With increasing viscosity, the peak in entropy remains constant. The DNS data has been obtained for base state viscosity values given in Table~\ref{tab: test_cases}.}
\label{fig: EntropyJumps}
\end{figure}

One dimensional governing equations of continuum gas dynamics (compressible Navier-Stokes) for an ideal gas are given by, 
\begin{align}
&\frac{\partial \rho^*}{\partial t^*} + \frac{\partial (\rho^* u^*) }{\partial x^*} = 0,\label{eq: NS_eqns1}
\\
&\frac{\partial}{\partial t^*}\left(\rho^* u^*\right)  + \frac{\partial}{\partial x^*}\left(\rho^* u^{*2}\right) = -\frac{\partial p^*}{\partial x^*} \nonumber \\
&+ \frac{\partial }{\partial x}\left(\left(\frac{4}{3}\mu^* + \mu^*_B\right)\frac{\partial u^*}{\partial x^*}\right),\label{eq: NS_eqns2}
\\
&\rho^* T^*\left(\frac{\partial  s^*}{\partial t^*} + u^*\frac{\partial s^*}{\partial x^*}\right) = \frac{\partial }{\partial x^*}\left( \frac{\mu^*C^*_p}{Pr} \frac{\partial T^*}{\partial x^*}\right) + \nonumber \\
&\left(\frac{4}{3}\mu^*+\mu^*_B\right)\left(\frac{\partial u^*}{\partial x^*}\right)^2,
\label{eq: NS_eqns3}
\end{align}
which are closed by the ideal gas equation of state, 
\begin{align}
p^* = \rho^* R^* T^*,\label{eq: IdealGas}
\end{align}
where $p^*, u^*, \rho^*, T^*, s^*$ respectively denote total pressure, velocity, density, temperature, and entropy of the fluid, $ x^*$ and $t^*$  denote space and time, and $\mu^*$ denotes dynamic viscosity. In this work, we perform DNS of Eqs.~\eqref{eq: NS_eqns1}-\eqref{eq: NS_eqns3} to resolve all the length scales of planar nonlinear acoustic waves. For our simulations (see Section~\ref{sec: numerics}), {\color{black}{we choose the gas specific constants for air at standard temperature and pressure (STP),
\begin{equation}
R^* = 287.105~\frac{\mathrm{m}^2}{\mathrm{s}^2\cdot K},\quad\quad\mu^*_B=0,\quad\quad Pr=0.72.
\end{equation}

\begin{figure}[!b]
\centering
\includegraphics[width=\linewidth]{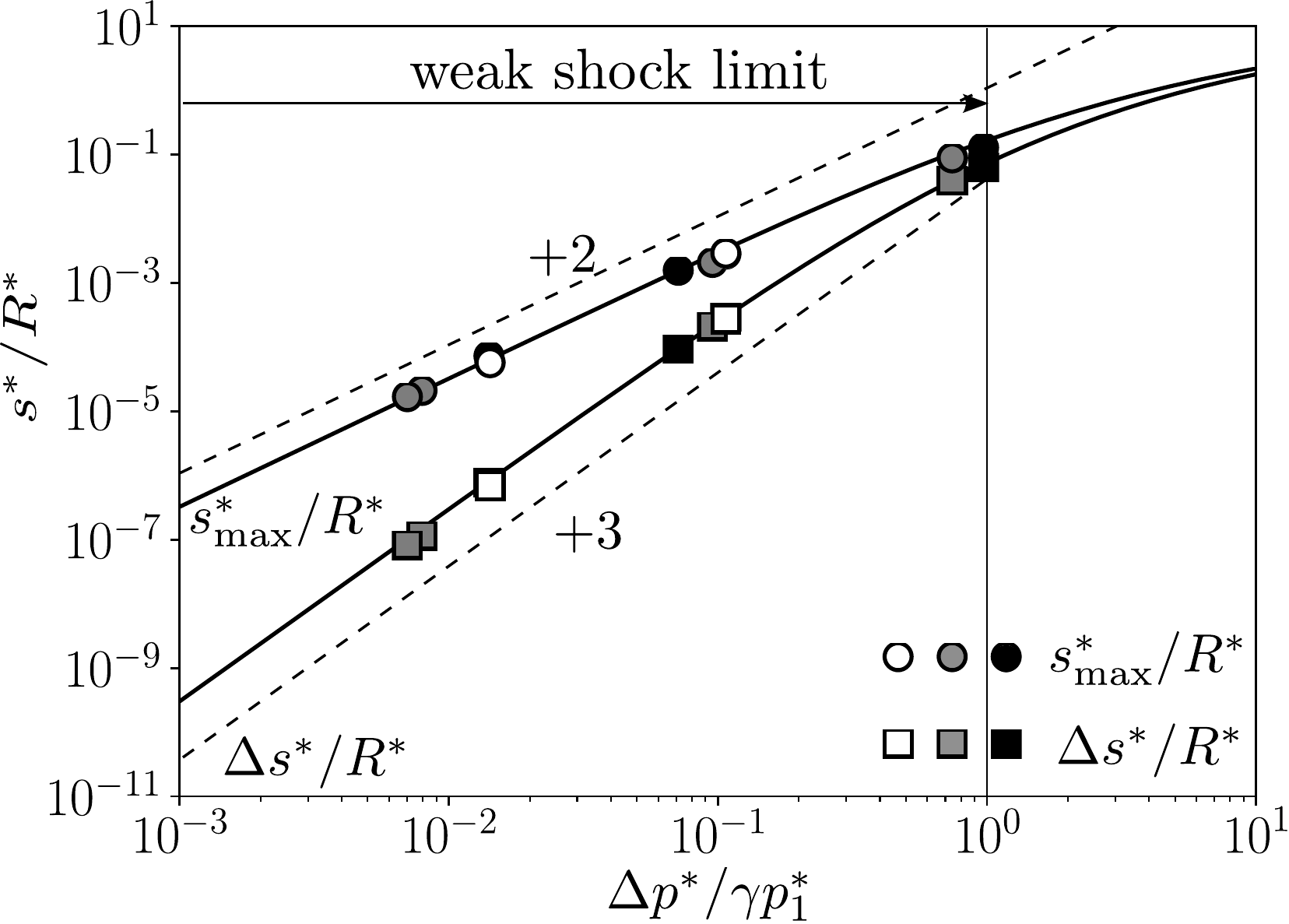}
\caption{{\color{black}{Entropy jump $\Delta s^*=s^*_2-s^*_1$ and maximum entropy generated $s^*_{\mathrm{max}}$ versus pressure jump $\Delta p^*$ across a planar shock wave. In the labeled region ($\Delta p^*/\gamma p^*_1 < 1$, referred as `weak shocks' hereafter), the entropy jump $\Delta s^*$ scales as $\mathcal{O}\left(\Delta p^{*3}\right)$, whereas the maximum entropy generated $s^*_{\mathrm{max}}$ scales as $\mathcal{O}\left(\Delta p^{*2}\right)$, approximately. Markers denote DNS data (see Section~\ref{sec: numerics}), (\protect\markerone,\protect\markerfour) $\mu^*=7.5\times10^{-3}~$kg$\cdot$m$^{-1}\cdot$s$^{-1}$; (\protect\markertwo,\protect\markerfive), $\mu^*=7.5\times10^{-4}~$kg$\cdot$m$^{-1}\cdot$s$^{-1}$; (\protect\markerthree,\protect\markersix), $\mu^*=7.5\times10^{-5}~$kg$\cdot$m$^{-1}\cdot$s$^{-1}$ for varying values of $\Delta p^*$. Solid lines correspond to Eqs.~\eqref{eq: EntropyMach} and~\eqref{eq: EntropyMax}.}}}
\label{fig: EntropyScaling}
\end{figure}
Planar nonlinear acoustic waves steepen and form weak shocks. \revision{For weak shocks, the smallest length scale (shock-thickness) is also significantly larger than the molecular length scales. Hence, in this work, we neglect the molecular vibrational effects in the single component ideal gas ($\mu^*_B=0$) , typically modeled via bulk viscosity effects~\cite{Cramer_PoF_2012}}. Across a freely propagating planar weak shock (Fig.\ref{fig: EntropyJumps}), the entropy jump ($\Delta s^* = s^*_2 - s^*_1$) is given by the classical gas-dynamic relation~\cite{LiepmannRoshko},
\begin{align}
\frac{\Delta s^*}{R^*} &= \frac{1}{\gamma - 1}\ln\left(1 + \frac{2\gamma}{\gamma + 1}\left(M^2 -1 \right)\right) \nonumber \\
&- \frac{\gamma}{\gamma - 1}\ln\left(\frac{\gamma + 1}{\gamma - 1 + 2/M^2}\right),
\label{eq: EntropyMach}
\end{align}
where $M$ is the Mach number, given by, 
\begin{equation}
 \frac{\Delta p^*}{\gamma p^*_1} = \frac{p^*_2 - p^*_1}{\gamma p^*_1} = \frac{2}{\gamma + 1}\left(M^2-1\right),
\label{eq: PressureMach}
\end{equation}
and $\Delta p^* = p^*_2 - p^*_1$ is the pressure jump with $p^*_1$ and $p^*_2$ being the pre-shock and post-shock pressures, respectively.}} Near the inflection point of the fluid velocity profile, the entropy reaches a local maximum $(s^* = s^*_\mathrm{max})$. According to Morduchow and Libby~\cite{Morris_JASc_1949}, maximum entropy $s^*_{\mathrm{max}}$ assuming $\mu^*_B = 0$ and $Pr = 3/4$, can be obtained as, 
\begin{equation}
\frac{s^*_{\mathrm{max}}}{R^*} = \frac{1}{\gamma - 1}\ln\left(1 + \frac{\gamma - 1}{2}M^2\left(1-\xi\right)\xi^{\frac{\gamma - 1}{2}}\right),
\label{eq: EntropyMax}
\end{equation}

where, 
\begin{equation}
\xi = \frac{\gamma - 1}{\gamma + 1} + \frac{2}{\gamma + 1}\frac{1}{M^2}.
\end{equation} 

For weak shock waves, ($\Delta p^*/\gamma p^*_1 <1$),  entropy jump $\Delta s^*$ and maximum entropy $s^*_{\mathrm{max}}$ scale with pressure jumps as (cf. Fig.~\ref{fig: EntropyScaling}),
\begin{equation}
{\Delta s^*} = \mathcal{O}\left(\Delta p^{*3}\right), \quad s^*_{\mathrm{max}} = \mathcal{O}\left({\Delta p}^{*2}\right),
\end{equation}
independent of $\mu^*$ (cf. Eqs.~\eqref{eq: EntropyMach} and~\eqref{eq: EntropyMax}). The overall entropy jump $\Delta s^*$ is due to irreversible thermoviscous losses occurring within the shocks. However, the overshoot in entropy ($s^*_{\mathrm{max}}>\Delta s^*$) is due to both reversible and irreversible processes, and is not in violation of the second law of thermodynamics~\cite{Morris_JASc_1949}. Moreover, in the range of pressure jumps considered in the DNS (see Section~\ref{sec: numerics}), the maximum Mach number of the shock is around $M\approx 1.4$, which is well within the limits of validity of the continuum approach~\cite{Bird_MolecularGasDynamicsBook_1994}. Hence, it is physically justified to draw conclusions regarding the smallest length scales through the governing equations based on continuum approach and assuming thermodynamic equilibrium.


\subsection{Perturbation variables and non-dimensionalization}
\label{sec: 2ndOrderEntropy}

{In this section, we utilize the previous consideration on the second order scaling of the maximum entropy $s^*_{\mathrm{max}}$ inside a weak shock wave to derive second order nonlinear acoustics equations. To this end, we decompose the variables in base state and perturbation fields and derive equations containing only linear and quadratic terms in perturbation fields. Denoting the base state with the superscript $( )_0$ and the perturbation fields with the superscript $({ })'$, we obtain, }
\begin{subequations}
\begin{align}
& \rho^* = \rho^{*}_0 + {\rho^*}',\quad p^* = p^*_0 + {p^*}', \\
& u^* = {u^*}', \quad s^* = {s^*}', \quad T^* = T^*_0 + {T^*}',
 \label{eq: PertDecom}
 \end{align}
\end{subequations}
where no mean flow $u^*_0 = 0$ is considered and $s^*_0$ is arbitrarily set to zero.
We neglect the fluctuations in the dynamic viscosity as well, i.e.,
\begin{equation}
\mu^* = \mu^*_0.
\end{equation}
{\color{black}{While in classic gas dynamics, pre-shock values are used to normalize fluctuations or jumps across the shock (e.g. see Eq.~\eqref{eq: PressureMach}), hereafter we choose base state values to non-dimensionalize the nonlinear acoustics equations,
\begin{subequations}
\begin{align}
& \rho = \frac{{\rho^*}}{\rho^*_0} = 1 + \rhoFluct ,~~p = \frac{p^*}{\gamma p^*_0} = \frac{1}{\gamma} + \pFluct, \\
& u = \frac{u^*}{a^*_0} = \uFluct, ~~s = \frac{s^*}{R^*} = \sFluct,~~ T = \frac{T^*}{T^*_0} = 1 + \TFluct,\\
&x = \frac{x^*}{L^*}, ~~ t = \frac{a^*_0 t^*}{L^*}.
\end{align}
\label{eq: per_norm}
\end{subequations}
where $L^*$ is the length of the one-dimensional periodic domain.  \revision{As also typically done in classical studies of homogeneous isotropic turbulence~\cite{Ishihara_ARFM_2009,Batchelor1953theory,pope2000turbulent,monin1971statistical,Hosokawa_PRE_2008, Burattini_PRE_2006}, periodic boundary conditions represent a common (yet not ideal) way to approximate infinite domains; as such, a spurious interaction between the flow physics that one wishes to isolate and the periodic box size may occur. For the TW and SW test cases analyzed herein, $L^*$ corresponds to the initial (and hence largest) reference length scale of the acoustic perturbation; in the AWT case, the value of $L^*$ should be chosen as much larger than the integral length scale $\ell$ or Taylor microscale $\lambda$ (see Section~\ref{sec: ScalesAcousticCascade}), which truly define the state of turbulence.}}



{Due to thermodynamic nonlinearities, wave propagation velocity increases across a high-amplitude compression front, resulting in wave-steepening ~\citep{Whitham2011} and hence generation of small length scales associated with increasing temperature and velocity gradients responsible for thermoviscous dissipation. Increase in thermoviscous dissipation results in positive entropy perturbations peaking within the shock structure. For pressure jumps $\Delta p^*/\gamma p^*_1<1$, the maximum entropy scales approximately as $\mathcal{O}\left(\Delta p^{*2}\right)$ (cf. Fig.~\ref{fig: EntropyScaling}). Moreover, as we discuss in the later section (see Section~\ref{sec: EnergyCorollary}), the second order nonlinear acoustic equations impose a strict limit of $|p'|<1/\gamma$ ($\simeq 0.714$ for $\gamma = 0.72$) for base state normalized (Eq.~\eqref{eq: per_norm}) (not pre-shock state normalized (Eq.~\eqref{eq: PressureMach})) perturbations. Hence, in our simulations~(see Section~\ref{sec: numerics}), we consider a suitable range of $10^{-3}<p'<10^{-1}$, which satisfies the aforementioned constraints. Thus, the second order scaling of entropy holds in our simulations. 

Below, we utilize this entropy scaling to derive the correct second order nonlinear acoustics equations governing the spatio-temporal evolution of dimensionless perturbation variables $p'$ and $u'$, as defined in Eq.~\eqref{eq: per_norm}.}

\subsection{Second order nonlinear acoustics equations}
\label{sec: 2ndOrderNLA}
For a thermally perfect gas, the differential in dimensionless density $\rho$ can be related to differentials in dimensionless pressure $p$ and dimensionless entropy $s$ as,
\begin{align}
d\rho &= \left(\frac{\partial \rho}{\partial p}\right)_{s} dp + \left(\frac{\partial \rho}{\partial s}\right)_{p} ds,\nonumber\\
& = \frac{\rho}{\gamma p}dp - \frac{\rho(\gamma - 1)}{\gamma}ds.
\label{eq: ConstitutiveEquation}
\end{align}
Nondimensionalizing the continuity Eq.~\eqref{eq: NS_eqns1} and substituting Eq.~\eqref{eq: ConstitutiveEquation}, we obtain, 
\begin{align}
&\frac{\partial \rho}{\partial t} + \frac{\partial \rho u}{\partial x} = 0,\nonumber\\
&\frac{\partial p}{\partial t} + u\frac{\partial p}{\partial x} + \gamma p\frac{\partial u}{\partial x} = (\gamma -1)p\left(\frac{\partial s}{\partial t} + u\frac{\partial s}{\partial x}\right).
\label{eq: PressureStep1}
\end{align}
Substituting the dimensionless forms of Eqs.~\eqref{eq: NS_eqns3} and~\eqref{eq: IdealGas} and utilizing the decomposition given in Eqs.~\eqref{eq: per_norm}, we obtain the following truncated equation for pressure perturbation $\pFluct$,
\begin{align}
\frac{\partial \pFluct}{\partial t} + \frac{\partial \uFluct}{\partial x} + \gamma \pFluct \frac{\partial \uFluct}{\partial x} &+ \uFluct\frac{\partial \pFluct}{\partial x} = \nu_0\left(\frac{\gamma - 1}{Pr}\right)\frac{\partial^2 \pFluct}{\partial x^2} \nonumber \\
& + \mathcal{O}\left(\pFluct\sFluct, \sFluct^2, \pFluct^3, \left(\frac{\partial \uFluct}{\partial x}\right)^2\right).\label{eq: pressure}
\end{align}
Similarly, the truncated equation for velocity perturbation $\uFluct$ is obtained as,
\begin{align}
\frac{\partial \uFluct}{\partial t} + \frac{\partial \pFluct}{\partial x} + \frac{\partial}{\partial x}\left(\frac{\uFluct^2}{2} - \frac{\pFluct^2}{2}\right) &= \frac{4}{3}\nu_0\frac{\partial^2 \uFluct}{\partial x^2} \nonumber \\
&+\mathcal{O}\left(\rho'^2\pFluct,\rho'^3\pFluct \right). \label{eq: velocity}
\end{align}
In Eqs.~\eqref{eq: pressure} and~\eqref{eq: velocity}, $\nu_0$ is the dimensionless kinematic viscosity given by, 
\begin{equation}
\nu_0 = \frac{\mu^*_0}{\rho^*_0 a^*_0L^*},
\label{eq: DimensionlessVisc}
\end{equation}
and quantifies viscous dissipation of waves relative to propagation. Equations~\eqref{eq: pressure} and~\eqref{eq: velocity} constitute the nonlinear acoustics equations truncated up to second order, governing spatio-temporal evolution of finite amplitude acoustic perturbations $\pFluct$ and $\uFluct$. The entropy scaling ($s^*_{\mathrm{max}}=\mathcal{O}\left(\Delta p^{*2}\right)$) discussed previously results in the dissipation term on the right hand side of Eq.~\eqref{eq: pressure}. Left hand side of Eqs.~\eqref{eq: pressure} and~\eqref{eq: velocity} contains terms denoting linear and nonlinear isentropic acoustic wave propagation. {Detailed derivation of Eqs.~\eqref{eq: pressure} and~\eqref{eq: velocity} is given in Appendix~\ref{sec: appedixA}, where we also show that the nonlinear terms on the left hand side (LHS) of Eqs.~\eqref{eq: pressure} and~\eqref{eq: velocity} are independent of the thermal equation of state. The functional form of the second order perturbation energy norm ($E^{(2)}$, Eq \eqref{eq: Energy_norm}) -- being exclusively dictated by such terms (see Section \ref{sec: EnergyCorollary}) -- is independent of the thermal equation of state of the gas. The results shown in this work focus on ideal-gas simulations merely for the sake of simplicity, with no loss of generality pertaining to inviscid nonlinear (up to second order) spectral energy transfer dynamics.}

  We note that, Eq.~\eqref{eq: pressure} consists of the velocity derivative term ($\gamma p'{\partial u'}/{\partial x}$), and is different from those obtained by Naugol'nykh and Rybak~\cite{Naugol1975spectrum}, which in dimensionless form read,
\begin{align}
&\frac{\partial \pFluct}{\partial t} - (\gamma -1)\pFluct\frac{\partial \pFluct}{\partial t} + \frac{\partial \uFluct}{\partial x} + \pFluct\frac{\partial \uFluct}{\partial x} = 0,\\
&\frac{\partial \uFluct}{\partial t} + \frac{\partial \pFluct}{\partial x} + \frac{\partial }{\partial x}\left(\frac{u^2}{2} - \frac{p^2}{2}\right) = 0.
\end{align} 
We adopt Eqs.~\eqref{eq: pressure} and~\eqref{eq: velocity} throughout the study since they represent the truncated governing equations exactly. Unlike Naugol'nykh and Rybak~\cite{Naugol1975spectrum}, we do not approximate the density $\rho$ using the Taylor series and only use the total differential form given in Eq.~\eqref{eq: ConstitutiveEquation}. 

Additionally, we note that Eqs.~\eqref{eq: pressure} and~\eqref{eq: velocity} can be combined into the Westervelt's equation~\cite{Hamilton_NLA_1998} only if the Lagrangian defined as, 
\begin{equation}
 \mathcal{L} = \frac{\uFluct ^2}{2} - \frac{\pFluct ^2}{2},
\end{equation}
is zero, which holds only for linear pure traveling waves. The derivation of the Burgers equation in nonlinear acoustics follows from the Westervelt's equation~\cite{Hamilton_NLA_1998}. Hence, it is inadequate in modeling general nonlinear acoustics phenomena involving mixed phasing of nonlinear waves which occurs in the Standing Wave (SW) and Acoustic Wave Turbulence (AWT) cases analysed here in.\section{Second order perturbation energy}
\label{sec: EnergyCorollary}
In this section we derive a new perturbation energy function for nonlinear acoustic waves utilizing Eqs.~\eqref{eq: pressure} and \eqref{eq: velocity}. To this end, we derive the perturbation energy conservation relation (energy corollary) for high amplitude acoustic perturbations. We show that the spatial average of the perturbation energy function satisfies the definition of the Lyapunov function for high amplitude acoustic perturbations and evolves monotonically in time (cf.~Fig.~\ref{fig: EnergyTime}). Utilizing the energy corollary, we derive spectral energy transport relations in further sections.

Multiplying Eqs.~\eqref{eq: pressure} and \eqref{eq: velocity} with $\pFluct$ and $\uFluct$ respectively and adding, we obtain,
\begin{align}
&\frac{\partial }{\partial t}\left(\frac{\pFluct ^2}{2} + \frac{\uFluct ^2}{2}\right) + \frac{\partial }{\partial x}\left(\uFluct\pFluct + \frac{\uFluct^3}{3}\right) + \gamma \pFluct ^2\frac{\partial \uFluct}{\partial x} = \nonumber \\ 
&\nu_0\left(\frac{\gamma - 1}{Pr}\right)\pFluct\frac{\partial^2 \pFluct}{\partial x^2} +   \frac{4}{3}\nu_0 u'\frac{\partial^2 \uFluct}{\partial x^2}. \quad\quad
\label{eq: EnergyCorollary_Open}
\end{align}
Spatial averaging of Eq.~\eqref{eq: EnergyCorollary_Open} over a periodic domain $[0, L]$ yields, 
\begin{align}
\frac{d\left\langle E^{(1)}\right\rangle}{dt} = -\left\langle \gamma p'^2\frac{\partial u'}{\partial x}\right\rangle &- \nu_0\left(\frac{\gamma - 1}{Pr}\right)\left\langle\left(\frac{\partial p'}{\partial x}\right)^2\right\rangle\nonumber\\
&-\frac{4}{3}\nu_0\left\langle\left(\frac{\partial u'}{\partial x}\right)^2\right\rangle,
\label{eq: LinearAcousticEnergy}
\end{align}
where $\left\langle . \right\rangle$ is the spatial averaging operator,
\begin{equation}
\left\langle . \right\rangle = \frac{1}{L}\int^L_0\left(.\right)dx,
\end{equation}
and 
\begin{equation}
E^{(1)} = \frac{\uFluct^2}{2} + \frac{\pFluct^2}{2},
\end{equation}
is the first order isentropic acoustic energy. Equation~\eqref{eq: LinearAcousticEnergy} suggests that, in a lossless medium ($\nu_0\rightarrow 0$), $\left\langle E^{(1)}\right\rangle$ would exhibit spurious non-monotonic behavior in time due to the first term on right hand side. Such non-monotonic behaviour is confirmed by the DNS results shown in Fig.~\ref{fig: EnergyTime}. Consequently, the linear acoustic energy norm $E^{(1)}$ does not quantify the perturbation energy correctly for high amplitude perturbations since the spatial average $\left\langle E^{(1)}\right\rangle$ supports spurious growth and decay in the absence of physical sources of energy.
\begin{figure}[!b]
\includegraphics[width=\linewidth]{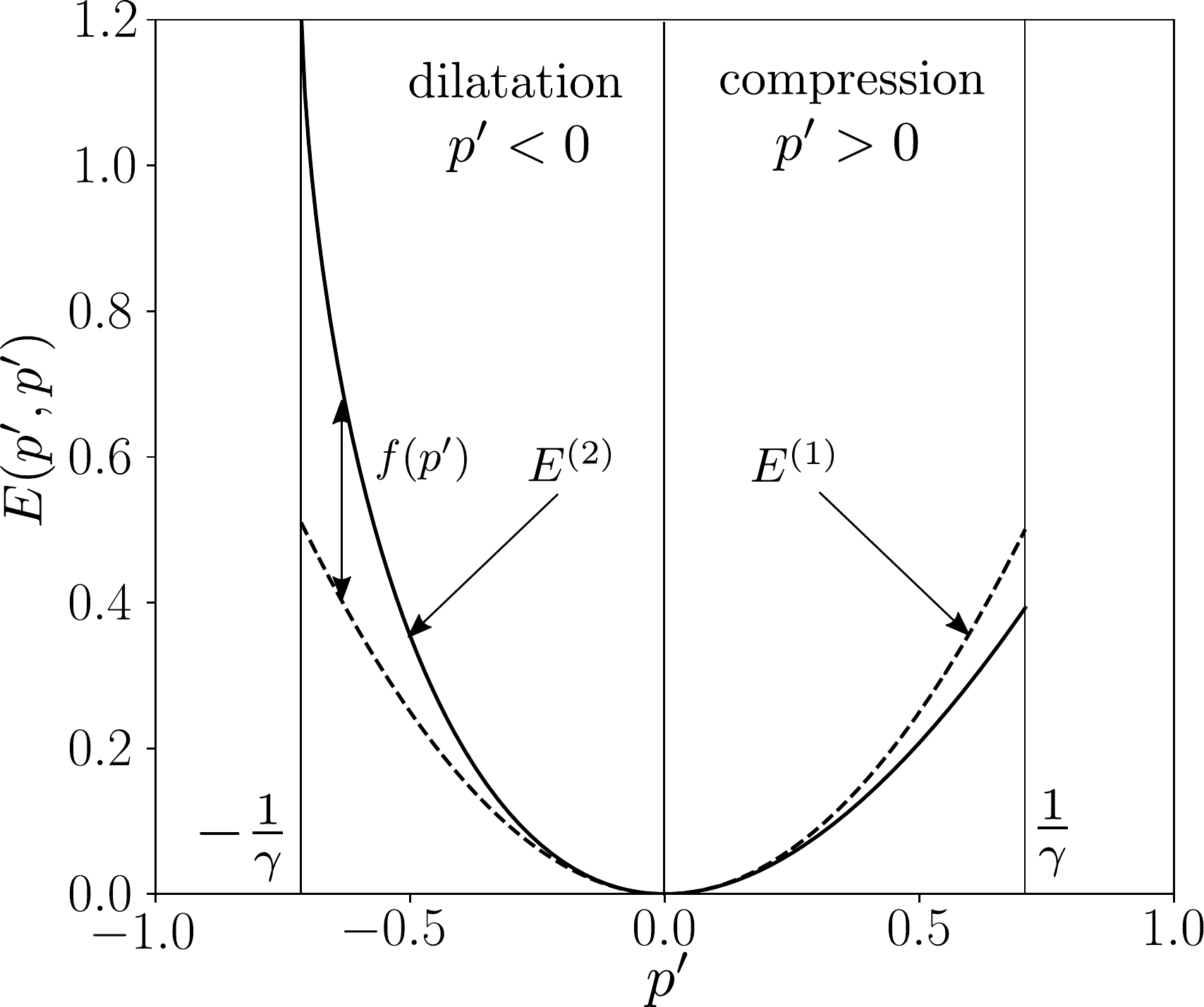}
\caption{Comparison of perturbation energy function for nonlinear acoustic waves $E^{(2)}$ with the linear acoustic energy $E^{(1)}$ in the case of $\pFluct=\uFluct$ (assumed for illustrative purpose). The correction $f(\pFluct)$ is independent of $\uFluct$.}
\label{fig: EnergyComp}
\end{figure}
The corrected perturbation energy function can be obtained upon recursively evaluating the velocity derivative term ($\gamma p'^2\partial u'/\partial x$ in Eq.~\eqref{eq: EnergyCorollary_Open}) utilizing Eq.~\eqref{eq: pressure} as, 
{\color{black}{
\begin{align}
\gamma p'^2\frac{\partial u'}{\partial x} &= -\frac{\partial }{\partial t}\left(\frac{\gamma \pFluct^3}{3}\right) - \frac{\partial}{\partial x}\left(\frac{\gamma \uFluct \pFluct^3 }{3}\right) -\nonumber \\ &\gamma\left(\gamma - \frac{1}{3}\right)\pFluct^3\frac{\partial \uFluct}{\partial x} 
- \frac{\nu_0(\gamma - 1)}{Pr}\gamma \pFluct^3\frac{\partial \pFluct}{\partial x}.
\label{eq: Rec1}
\end{align}
Furthermore, the third term in above Eq.~\eqref{eq: Rec1} on the right can be further evaluated as,
\begin{align}
&\gamma\left(\gamma - \frac{1}{3}\right)\pFluct^3\frac{\partial \uFluct}{\partial x} = -\frac{\partial }{\partial t}\left(\frac{\gamma}{4}\left(\gamma - \frac{1}{3}\right)\pFluct^4\right) \nonumber \\
&- \frac{\partial}{\partial x}\left(\frac{\gamma}{4}\left(\gamma - \frac{1}{3}\right)\uFluct\pFluct^4\right) - \gamma\left(\gamma - \frac{1}{3}\right)\left(\gamma - \frac{1}{4}\right)\pFluct^4\frac{\partial \uFluct}{\partial x}\nonumber \\
&-\frac{\nu_0(\gamma -1 )}{Pr}\gamma\left(\gamma - \frac{1}{3}\right)\pFluct^3\frac{\partial^2 \pFluct}{\partial x^2},
\label{eq: Rec2}
\end{align}
and so on}}. Continued substitution according to Eqs.~\eqref{eq: Rec1} and~\eqref{eq: Rec2} yields the closure of the system and the following energy corollary,
\begin{equation}
\frac{\partial E^{(2)}}{\partial t} + \frac{\partial I}{\partial x}= \nu_0\left(\frac{\gamma - 1}{Pr}\right) h(\pFluct)\frac{\partial^2 \pFluct}{\partial x^2} +  \frac{4}{3}\nu_0 \uFluct\frac{\partial^2 \uFluct}{\partial x^2} ,
\label{eq: energy_cons}
\end{equation}
where,  
\begin{equation}
I(\pFluct,\uFluct) = \pFluct \uFluct + \frac{\uFluct^3}{3} + \uFluct f(\pFluct),
\label{eq: Intensity}
\end{equation} 
is the intensity (energy flux) of the field, $h(\pFluct)$ is given by,
\begin{equation}
h(\pFluct) = \pFluct + \frac{\partial f(\pFluct)}{\partial \pFluct} = \frac{\partial E^{(2)}}{\partial \pFluct}.
\end{equation}
and $E^{(2)}$ is given by,
\begin{equation}
E^{(2)}(\pFluct,\uFluct) =  \frac{\uFluct ^2}{2} + \frac{\pFluct ^2}{2} + f(\pFluct) = E^{(1)} + f(\pFluct),
\label{eq: Energy_norm}
\end{equation}
and defines the second order perturbation energy for high amplitude acoustic perturbations. The energy corollary Eq.~\eqref{eq: energy_cons} is mathematically exact for the governing Eqs.~\eqref{eq: pressure} and~\eqref{eq: velocity}.

\begin{figure*}[!t]
\centering
\includegraphics[width=\textwidth]{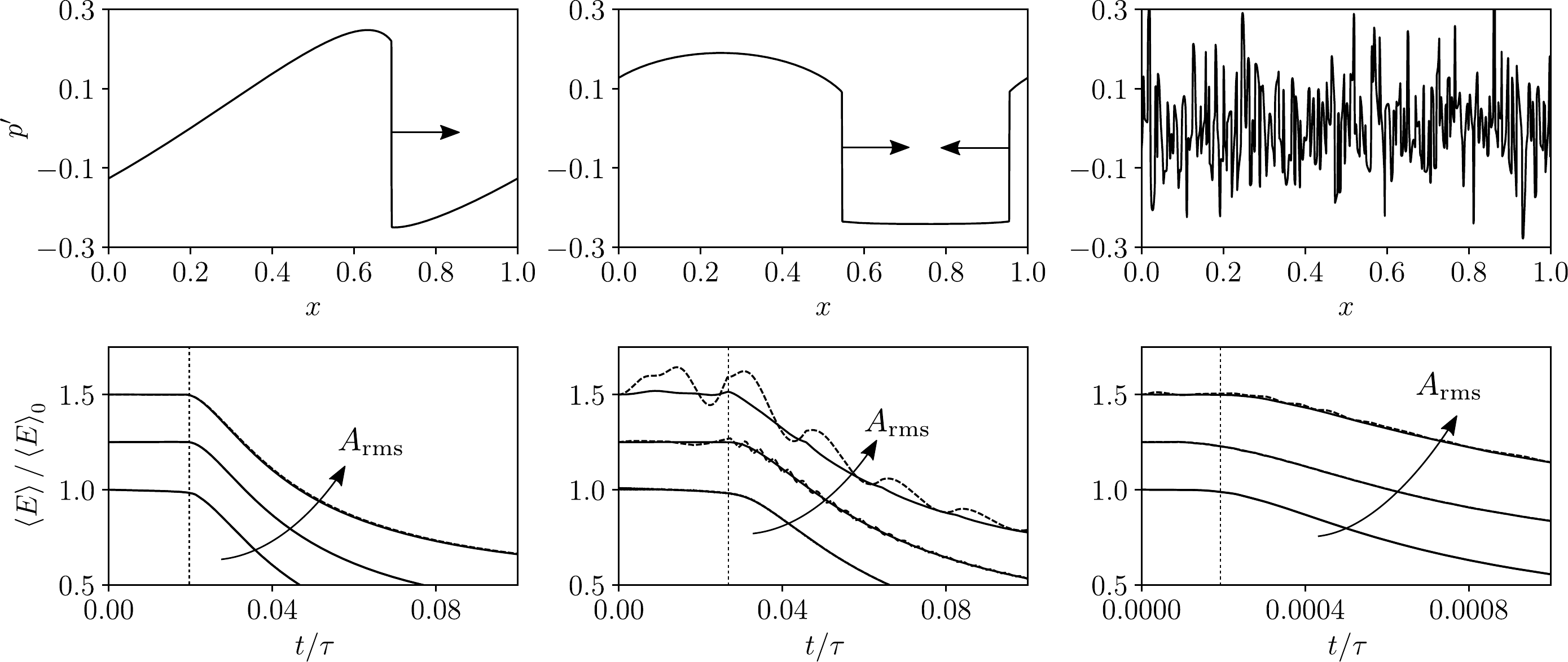}
\put(-510,220){$(a)$}
\put(-330,220){$(b)$}
\put(-160,220){$(c)$}
\caption{Spatial profile of finite amplitude waves (top) for TW($a$), SW($b$), and AWT($c$). Evolution of the average perturbation energy (< $E^{(2)}$ > (--); < $E^{(1)}$ > ($--$)) evaluated from the DNS data (bottom) scaled by the initial value against scaled time $t/\tau$ (cf. Eq.~\eqref{eq: SteepeningTime}) for increasing values of perturbation amplitude $A_{\mathrm{rms}}$ defined in Eq.~\eqref{eq: PcDef} at $\nu_0 = $ 1.836$\times$10$^{-7}$ (see Table~\ref{tab: test_cases}). The curves are shifted vertically by 0.25 for illustrative purpose only. With increasing perturbation amplitude $A_{\mathrm{rms}}$, the variation of linear acoustic energy norm < $E^{(1)}$ > becomes increasingly non-monotonic. The vertical dashed line (bottom) highlights the end of approximately inviscid spectral energy cascade regime. In this regime, the energy is primarily redistributed in the spectral space due to the nonlinear propagation ($\epsilon\simeq 0$).}
\label{fig: EnergyTime}
\end{figure*}

The correction term $f(\pFluct)$ in $E^{(2)}$ appears due the thermodynamic nonlinearities and can be derived in the closed form as,
\begin{equation}
f(\pFluct) = \sum^{\infty}_{n=2}T_n =  \sum^{\infty}_{n=2}(-1)^{n+1}\frac{\gamma \pFluct^{n+1}}{n+1}\prod^{n}_{i=3}\left(\gamma - \frac{1}{i}\right),
\label{eq: EnergyCorrection_app}
\end{equation}
{where $T_2$ and $T_3$ can be identified in Eqs.~\eqref{eq: Rec1} and \eqref{eq: Rec2}, respectively.} Isolating the $n^{th}$ term of the above infinite series as,
\begin{equation}
T_n = (-1)^{n+1}\frac{\gamma \pFluct^{n+1}}{n+1}\underbrace{\left(\gamma - \frac{1}{3}\right)\left(\gamma - \frac{1}{4}\right)\cdots\left(\gamma - \frac{1}{n}\right)}_{n-2~\mathrm{terms}}.
\label{eq: nth-term}
\end{equation}
Multiplied fractions in the Eq.~\eqref{eq: nth-term} above yield the $n^{th}$ term as,
\begin{equation}
T_n = -\frac{2\gamma}{\left(\gamma-1\right)\left(2\gamma - 1\right)}\left(\gamma \pFluct\right)^{n+1}\binom{1/\gamma}{n+1}.
\end{equation}
Finally, the energy correction $f(\pFluct)$ can be recast as, 
\begin{align}
f(\pFluct) = \sum^{\infty}_{n=2}T_n = &-\frac{2\gamma}{\left(\gamma-1\right)\left(2\gamma - 1\right)}\Big(\left(1+\gamma \pFluct\right)^{1/\gamma} - 1 - \nonumber \\
& \pFluct+ \frac{\left(\gamma - 1\right)\pFluct^2}{2}\Big).
 \label{eq: EnergyCorrection}
\end{align}
The correction function $f(\pFluct)$ defined in Eq.~\eqref{eq: EnergyCorrection} accounts for second order isentropic nonlinearities and is not a function of entropy perturbation. Hence, $E^{(2)}$ accounts for the effect of high 
amplitude perturbations on perturbation energy isentropically. We note that this separates $E^{(2)}$ fundamentally from generalized linear perturbation energy norms, such as the ones derived by Chu~\cite{Chu1965EnergyNorm} for small amplitude non-isentropic perturbations, and by Meyers~\cite{myers1986exact} for acoustic wave propagation in a steady flow. {Moreover, as discussed in the previous section (and shown in Appendix~\ref{sec: appedixA}), since the isentropic nonlinearities on the LHS of Eqs.~\eqref{eq: pressure} and~\eqref{eq: velocity} are independent of the thermal equation of state, the functional form of $E^{(2)}$ and $I$ are also independent of the equation of state. However, the dissipation term on the right hand side of the energy corollary Eq.~\eqref{eq: energy_cons} may change with the thermal equation of state.}

The energy correction $f(\pFluct)$ {is infinite order in pressure perturbation $\pFluct$} and converges only for perturbation magnitude $|\pFluct|<1/\gamma$ thus naturally yielding the strict limit of validity of second order acoustic equations in modelling wave propagation and wave steepening. Figure~\ref{fig: EnergyComp} shows the newly derived second order perturbation energy $E^{(2)}$ compared against the isentropic acoustic energy $E^{(1)}$. Both $E^{(2)}$ and $E^{(1)}$ are non-negative in the range $|\pFluct|<1/\gamma$ ($\pFluct=\uFluct$ is assumed for illustrative purpose). Furthermore, $E^{(2)}$ is asymmetric in nature, with larger energy in dilatations compared to compressions of same magnitude, as shown in Fig.~\ref{fig: EnergyComp}. Such asymmetry signifies that the medium (\revision{compressible ideal gas} in the present study) relaxes towards the base state faster for finite dilatations compared to compressions. 

For compact supported or spatially periodic perturbations, the energy conservation Eq.~\eqref{eq: energy_cons} shows that the spatially averaged energy $\left\langle E^{(2)}\right\rangle$ decays monotonically in time (in the absence of energy sources) accounting for the nonlinear interactions i.e.,
\begin{align}
\dot{\mathscr{V}}=\frac{d \left\langle E^{(2)}\right\rangle}{dt} &= -\nu_0\left(\frac{\gamma - 1}{{Pr}}\right)\left\langle \frac{\partial^2 E^{(2)}}{\partial \pFluct^2}\left(\frac{\partial \pFluct}{\partial x}\right)^2\right\rangle \nonumber \\
&-  \frac{4}{3}\nu_0\left\langle\left(\frac{\partial \uFluct}{\partial x}\right)^2\right\rangle \nonumber \\
&= \left\langle \mathcal{D}\right\rangle= -\epsilon\leq 0,
\label{eq: LyapunovFunction}
\end{align}
where $\mathcal{D}$ is the perturbation energy dissipation and $\epsilon$ is the negative of its spatial average. The spatial average $\left\langle E^{(2)}\right\rangle$ is non-negative ($E^{(2)}\geq 0$), and Eq.~\eqref{eq: LyapunovFunction} and Fig.~\ref{fig: EnergyTime} confirm that $\left\langle E^{(2)}\right\rangle$ evolves monotonically in time in the absence of physical energy sources. Hence, the spatial average of the perturbation energy function $\left\langle E^{(2)}\right\rangle$ defines the Lyapunov function $\mathscr{V}$ of the nonlinear acoustic system governed by the set of second order governing Eqs.~\eqref{eq: pressure} and \eqref{eq: velocity} exactly. The spatial average $\left\langle E^{(2)}\right\rangle$ should be used for studying the stability of nonlinear acoustic systems~\cite{George2012, Strogatz2018NonlinearDynamics}, which, however, falls beyond the scope of this work.

Wave-front steepening entails cascade of perturbation energy into higher wavenumbers thus broadening the energy spectrum. Fully broadened spectrum of acoustic perturbations exhibits energy at very small length scales which causes high thermoviscous energy dissipation. We analyse the separation of length scales and energy decay caused by nonlinear wave steepening and thermoviscous energy dissipation in the following sections. To this end, we utilize the direct numerical integration of Navier-Stokes Eqs.~\eqref{eq: NS_eqns1}-\eqref{eq: IdealGas} resolving all the length scales (DNS) and the exact energy corollary~Eq.~\eqref{eq: energy_cons} for second order truncated Eqs.~\eqref{eq: pressure} and~\eqref{eq: velocity}.


\section{High Fidelity Simulations with Adaptive Mesh Refinement}
\label{sec: numerics}
We perform shock-resolved numerical simulations of 1D Navier-Stokes (DNS) Eqs.~\eqref{eq: NS_eqns1}-\eqref{eq: IdealGas} with Adaptive Mesh Refinement (AMR). We use the perturbation energy $E^{(2)}$ defined in Eq.~\eqref{eq: Energy_norm} to define the characteristic dimensionless perturbation amplitude $A_{\mathrm{rms}}$ as, 
\begin{equation}
 A_{\mathrm{rms}} = \sqrt{\left\langle E^{(2)}\right\rangle},
\label{eq: PcDef}
\end{equation}
which is varied in the range $10^{-3}-10^{-1}$. The dimensionless kinematic viscosity at base state $\nu_0$ is also varied from 1.836$\times$10$^{-5}$ to 1.836$\times$10$^{-7}$. The base state conditions in the numerical simulations correspond to STP, i.e. $p^*_0 = 101325~\mathrm{Pa}$ and $T^*_0 = 300~\mathrm{K}$.
\begin{table}[!t]
\def\arraystretch{2}
\begin{ruledtabular}
\begin{tabular}{c c c c}
$\nu_0$	&  1.836$\times$10$^{-5}$ &  1.836$\times$10$^{-6}$ &  1.836$\times$10$^{-7}$\\
\hline
$A_{\mathrm{rms},0}$ 	& $10^{-3}$ & $10^{-2}$ & $10^{-1}$ \\
\hline
\hline
$u^*_{\mathrm{rms}}~\left(\mathrm{m}/\mathrm{s}\right)$	&  0.347 &  3.472&  34.725\\
\hline
$p^*_{\mathrm{rms}}~\left(\mathrm{kPa}\right)$	&  0.142 &  1.419 &  14.185\\
\end{tabular}
\end{ruledtabular}
\caption{Simulation parameter space for TW, SW, and AWT cases listing base state dimensionless viscosity $\nu_0$ (cf. Eq.~\eqref{eq: DimensionlessVisc}), initial characteristic perturbation amplitude $A_{\mathrm{rms},0}$ (cf. Eq.~\eqref{eq: PcDef}), and dimensional characteristic perturbation in velocity $u^*_{\mathrm{rms}}$ and pressure $p^*_{\mathrm{rms}}$ fields (Eq.~\eqref{eq: AcousticReynolds}).}
 \label{tab: test_cases}
\end{table}

The goal of spanning $A_{\mathrm{rms}}$ and $\nu_0$ over three orders of magnitude is to achieve widest possible range of energy cascade rate and dissipation within computationally feasible times. Equation~\eqref{eq: PcDef} yields the definitions of the perturbation Reynolds number~$\mathrm{Re}_L$ , characteristic perturbation velocity field~$u^*_{\mathrm{rms}}$, and pressure field $p^*_{\mathrm{rms}}$ as, 
\begin{equation}
\mathrm{Re}_L = \frac{A_{\mathrm{rms}}a^*_0 L^*}{\nu_0^*},~\quad u^*_{\mathrm{rms}} = a^*_0A_{\mathrm{rms}},~\quad p^*_{\mathrm{rms}} = \rho^*_0{a_0^*}^2A_{\mathrm{rms}},
\label{eq: AcousticReynolds}
\end{equation}
where $\mathrm{Re}_L$ denotes ratio of diffusive to wave steepening time scale over the length $L$.
{In the simulations, we keep $\mathrm{Re}_L\gg 1$, which corresponds to very fast wave steepening rates compared to diffusion.} In further sections (see Section~\ref{sec: ScalesAcousticCascade}), we define the wave turbulence Reynolds number $\mathrm{Re}_{\ell}$ based on the integral length scale $\ell$. Below, we briefly discuss the numerical scheme utilized for shock-resolved simulations and outline the initialization of the three configurations (TW, SW, and AWT) for numerical simulations.

\subsection{Numerical approach}
\begin{figure}[!t]
\centering
\includegraphics[width=\linewidth]{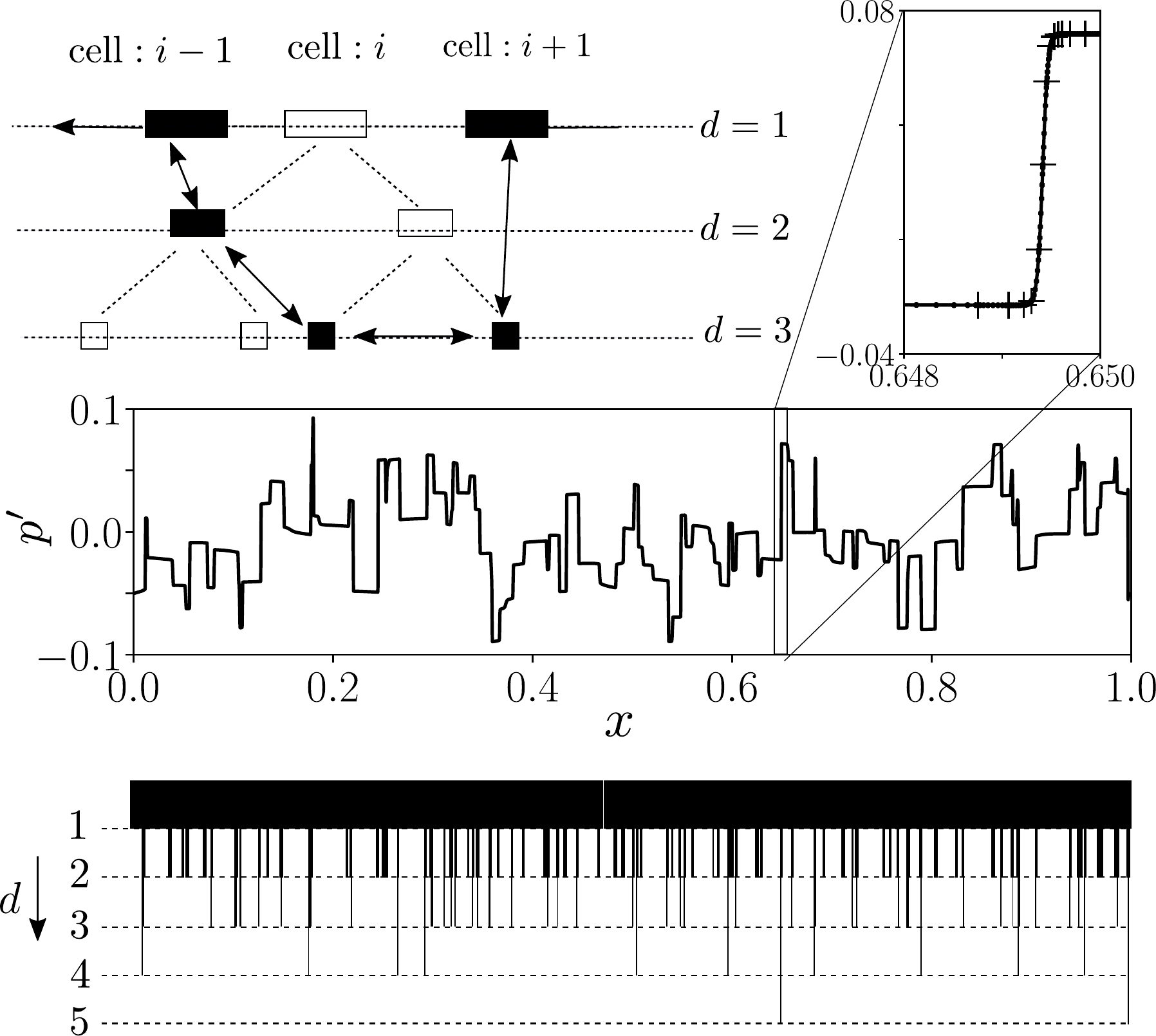}
\caption{Illustration of the binary tree implementation of Adaptive Mesh Refinement (AMR) technique (top left). The mesh is refined based on the resolution error in pressure field in each cell acting as a node of a binary tree. The pressure field shown (middle) corresponds to the randomly initialized AWT case with $A_{\mathrm{rms}}=10^{-1}$, $\nu_0 =  1.836\times$10$^{-6}$ (Table~\ref{tab: test_cases}) at $t/\tau = 0.04$. The inset shows the resolved shock wave with $(+)$ denoting the cell interfaces. The mesh refinement levels (bottom) show the depth $d$ of the binary tree.}
\label{fig: AMR}
\end{figure}
We integrate the fully compressible 1D Navier-Stokes Eqs.~\eqref{eq: NS_eqns1}-\eqref{eq: NS_eqns3} in time utilizing the staggered spectral difference (SD) spatial discretization approach~\cite{kopriva1996conservative}. In the SD approach, the domain is discretized into 
cells. Within each cell, the orthogonal polynomial reconstruction of variables allows numerical differentiation with spectral accuracy. We refer the reader to the work by Kopriva and Kolias~\cite{kopriva1996conservative} for further details.

To accurately resolve spectral energy dynamics at all length scales, i.e. for resolved weak shock waves, we combine the SD approach with the adaptive mesh refinement (AMR) approach as first introduced by Mavriplis~\cite{mavriplis1994adaptive} for spectral methods. The SD-AMR approach eliminates the computational need of very fine grid everywhere for resolving the propagating shock waves. To this end, we expand the values of a generic variable $\phi$ local to the cell in the Legendre polynomial space as,
\begin{equation}
\phi = \sum^{N}_{i=1}\hat{\phi}_i \psi_i(x)
\end{equation}
where $\psi_i(x)$ is the Legendre polynomial of $(i-1)^{\mathrm{th}}$ degree. The polynomial coefficients $\hat{\phi}_i$ are utilized for estimating the local resolution error $\varepsilon$ defined as~\cite{mavriplis1994adaptive},
\begin{equation}
\varepsilon = \left(\frac{2 \hat{\phi}^2_N}{2N + 1} + \int^{\infty}_{N+1}\frac{2 f_{\varepsilon}^2(n)}{2n + 1} dn\right)^{1/2}, \quad f_{\varepsilon}(n) = ce^{-\sigma n},
\end{equation} 
where $f_{\varepsilon}$ is the exponential fit through the coefficients of the last four modes in the Legendre polynomial space.
As the estimated resolution error $\varepsilon$ exceeds a pre-defined tolerance, the cell divides into two subcells, which are connected utilizing a binary tree (shown in Fig.~\ref{fig: AMR}). The subcells merge together if the resolution error decreases below a pre-defined limit.

\subsection{Initial conditions}
\begin{table}[!b]
\setlength{\tabcolsep}{0.95em}
\def\arraystretch{1.2}
\begin{ruledtabular}
\begin{tabular}{c c c c}
{$~~$} 	& TW & SW & {AWT}\\
\hline
$k_0$	&  1  &  1 &  1 \\
$k_E$	&  1  &  1 &  100 \\
$b_0(k)$	&  0  &  0 &  $e^{-(|k|-k_E)^2}$ \\
{$\widehat{E}_k$} & {${A}_{\mathrm{rms}}^2\delta(k_0)$} & {${A}_{\mathrm{rms}}^2\delta(k_0)$} & ${A}_{\mathrm{rms}}^2$\\
\end{tabular}\quad
\end{ruledtabular}
 \caption{Initial spectral compositions for traveling wave (TW), standing wave (SW), and acoustic wave turbulence (AWT). $\delta\left(\cdot\right)$ is the Dirac delta function.}
 \label{tab: InitialSpectra}
\end{table}
We utilize the Riemann invariants for compressible Euler equations to initialize the propagating traveling and standing wave cases in the numerical simulations. The Riemann invariants in terms of perturbation variables assuming nonlinear isentropic changes are given by, 
\begin{align}
 R_{-} = \frac{2}{\gamma -1 }\left(\left(1 + {\rhoFluct}\right)^{\frac{\gamma - 1}{2}} - 1\right) - \uFluct,\label{eq: InvariantL} \\
 R_{+} = \frac{2}{\gamma -1 }\left(\left(1 + {\rhoFluct}\right)^{\frac{\gamma - 1}{2}} - 1\right) + \uFluct,
\label{eq: InvariantR}
\end{align}
where $R_{-}$ and $R_{+}$ are the left and right propagating invariants, respectively, and $\uFluct$ and ${\rhoFluct}$ are normalized velocity and density perturbations, as defined in Eq.~\eqref{eq: per_norm}. Initial conditions for TW and SW cases correspond to $R_{-} = 0$ and $R_{-} = R_{+}$ respectively. 

To initialize the broadband noise case, we first choose $\pFluct$ and $\uFluct$ pseudo-randomly from a uniform distribution for the whole set of discretization points in $x$. Low-pass filtering of $\pFluct$ and $\uFluct$ yields, 
\begin{align}
 \widetilde{\widehat{p}_k}(t&=0) = \widehat{p}_kb_0(k),\quad \widetilde{\widehat{u}_k}(t=0) = \widehat{u}_kb_0(k)\nonumber\\
  b_0(k) &= \begin{cases} 
      1 & k_0\leq|k|\leq k_E \\
     e^{-(|k|-k_E)^2} & |k|> k_E
   \end{cases}.
\label{eq: FilterPU1}
\end{align}
where ${\widehat{p}_k}$ and ${\widehat{p}_k}$ are the Fourier coefficients of pseudo-random fields $\pFluct$ and $\uFluct$, respectively. $\widetilde{\widehat{p}_k}$ and $\widetilde{\widehat{u}_k}$ are the low-pass filtered coefficients. Inverse Fourier transform of Eq.~\eqref{eq: FilterPU1} yields smooth initial conditions with the initial spectral energy ${\widehat{E}_k}$, as defined in Section~\ref{sec: ScalesAcousticCascade} (cf. Eq.~\eqref{eq: spectral_energy}). For TW and SW, only single harmonic ($k=1$ in the current work) contains all of the initial energy. However, for AWT, $\widehat{E}_k$ is governed by the correlation function of velocity and pressure fields. In Table~\ref{tab: InitialSpectra}, we summarize the initial spectral energy for all three cases based on Eq.~\eqref{eq: FilterPU1}.\section{Scales of acoustic energy cascade and dissipation}
\label{sec: ScalesAcousticCascade}

In this section, we derive the analytical expressions of spectral energy, energy cascade flux, and spectral energy dissipation utilizing the exact energy corollary Eq.~\eqref{eq: energy_cons} (see Section~\ref{sec: EnergyCorollary}). We then identify the integral length scale $\ell$, the Taylor microscale $\lambda$, and the Kolmogorov length scale $\eta$ for TW, SW, and AWT cases in a periodic domain utilizing the DNS data (see Fig.~\ref{fig: LengthScalesSummary} and Table~\ref{tab: LengthScalesTable}). Temporal evolution laws of these length scales yield energy decay laws, which are used for dimensionless spectral scaling relations (see Section~\ref{sec: SpectralScaling}).

\subsection{Spectral energy flux and dissipation rate for periodic perturbations}

The exact perturbation energy conservation equation is given by (cf. Eq.~\eqref{eq: energy_cons}),
\begin{equation}
\frac{\partial E^{(2)}}{\partial t} + \frac{\partial I}{\partial x} = \mathcal{D}
\end{equation}
Integrating over the periodic domain, the above energy corollary can be converted into the following statement of conservation of perturbation energy in the spectral space,
\begin{equation}
\frac{d}{d t} \sum_{|k'|\leq k} \widehat{E}_{k'} + \widehat{\Pi}_k = \sum_{|k'|\leq k}\widehat{\mathcal{D}}_{k'},
\label{eq: spectral_conservation}
\end{equation}
\begin{table}[!t]
\setlength{\tabcolsep}{0.2em}
\def\arraystretch{1.2}
\begin{ruledtabular}
\begin{tabular}{c c c c}
\multirow{2}{*}{Length scale} 	& Integral & Taylor & Kolmogorov\\
                            	& length scale & Microscale & length scale\\
                                &$\ell$  & $\lambda$ & $\eta$\\
\hline
\\
{Definition}	& $\sqrt{\frac{\sum_k {\widehat{E}_k}/{k^2}}{\sum_k \widehat{E}_k}}$  &  $\sqrt{\frac{2\delta\left\langle E^{(2)}\right\rangle }{\epsilon}} $ &  $\frac{\delta}{\sqrt{\left\langle E^{(2)}\right\rangle}}$ \\
Characteristic	&  \multirow{2}{*}{$(k_0,k_E)$} &  \multirow{2}{*}{$(k_E, k_{\delta})$} &  \multirow{2}{*}{$(k_\delta, \infty)$} \\
spectral range	&     &    &      \\
\end{tabular}\quad
\end{ruledtabular}
 \caption{ Summary of the three length scales $\ell$, $\lambda$, and $\eta$, respective definitions, and the range of spectrum characterized by them. The integral length scale characterizes the energy containing range $(k_0, k_E)$. The Taylor microscale is the characteristic of the energy transfer and dissipation range $(k_E, k_{\delta})$. The Kolmogorov length scale corresponds to the highest wavenumber generated as a result of nonlinear acoustic energy cascade.}
 \label{tab: LengthScalesTable}
\end{table}
\begin{figure}[!t]
\centering
\includegraphics[width=\linewidth]{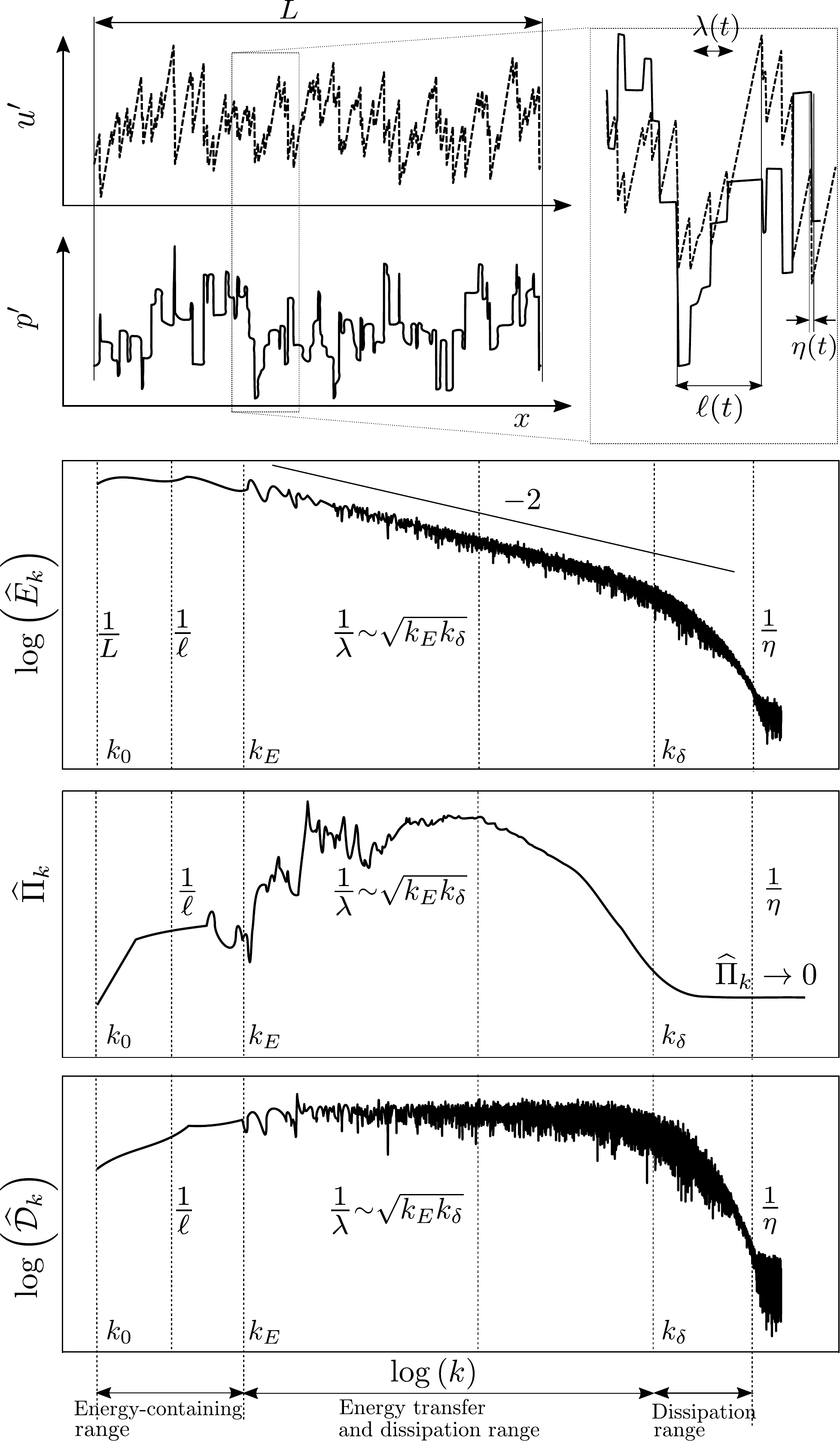}
\put(-245,410){$(a)$}
\put(-245,295){$(b)$}
\put(-245,195){$(c)$}
\put(-245,110){$(d)$}
\caption{ {Schematic illustrating the global picture of various length scales associated with spectral energy cascade in nonlinear acoustics in both spatial $(a)$ and spectral $(b)-(d)$ space. $(a)$ shows the perturbation velocity $\uFluct$ ($--$) and pressure $\pFluct$ (--) fields in AWT obtained from the DNS data for $\nu_0 = $1.836$\times$10$^{-7}$ and $A_{\mathrm{rms}}=10^{-1}$. $(b)$ shows the corresponding spectral energy $\widehat{E}_k$ in log-log space. The spectral flux $\widehat{\Pi}_k$ and dissipation $\widehat{\mathcal{D}}_k$ are shown in $(c)$ and $(d)$ respectively. The integral length scale $\ell$ corresponds to the characteristic distance between the shock waves traveling in the same direction. The Kolmogorov length scale $\eta$ corresponds to the shock wave thickness. The Taylor microscale $\lambda$ is the diffusive length scale and satisfies $\ell\gg\lambda\gg\eta$. $L$ corresponds to the length of the domain.}}
\label{fig: LengthScalesSummary}
\end{figure}
\begin{figure*}[!t]
\centering
\includegraphics[width=\textwidth]{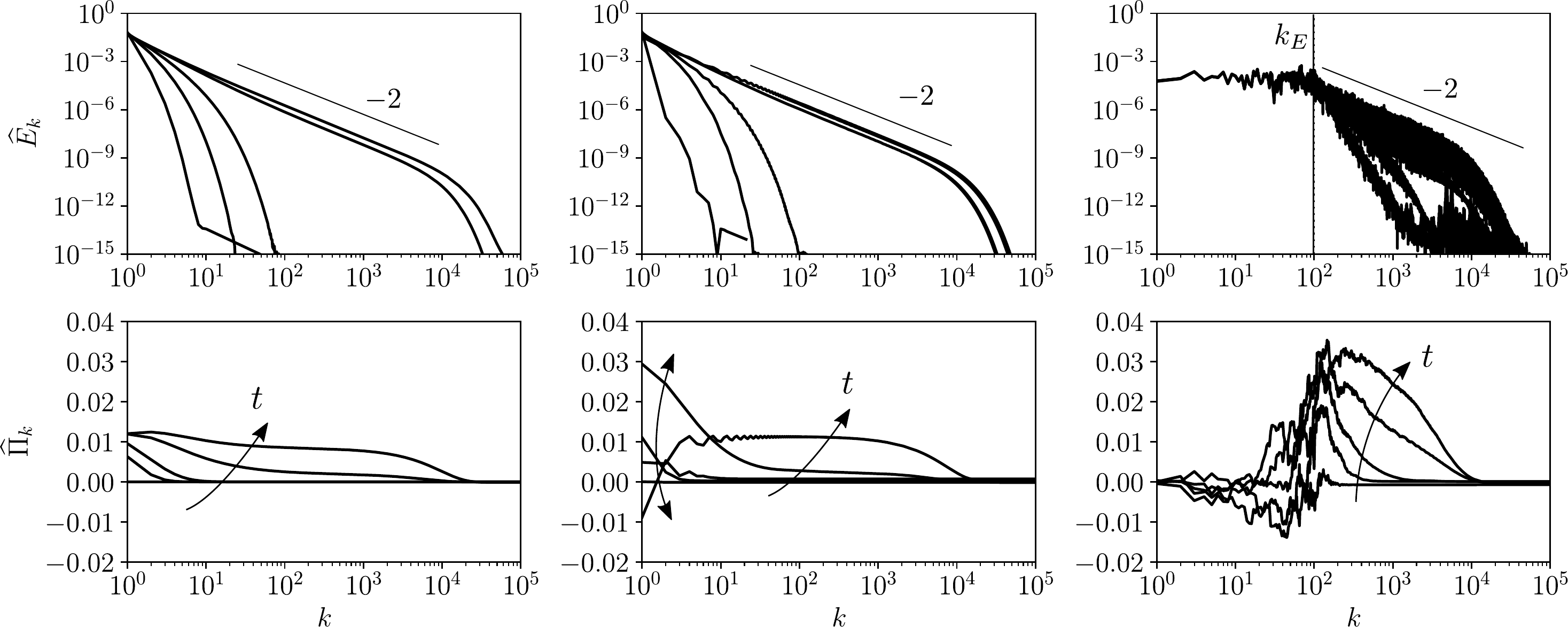}
\put(-510,210){$(a)$}
\put(-330,210){$(b)$}
\put(-160,210){$(c)$}
\caption{Spectro-temporal evolution of $\widehat{E}_k$ (top) and spectral flux $\widehat{\Pi}_k$ (bottom) for a TW ($a$), SW ($b$) and AWT ($c$). Spectral flux $\widehat{\Pi}_k$ for a traveling wave simply increases towards high wavenumbers. For a standing wave, $\widehat{\Pi}_k$ oscillates at low wavenumbers cyclically due to collisions of oppositely traveling shock waves while high wavenumber behaviour resembles that of a traveling shock. For AWT, the spectral broadening occurs for $k>k_E$ {with small fluctuations in time for $k<k_E$}.}
\label{fig: EnergyFLux}
\end{figure*}
where the first term corresponds to the temporal rate of change of cumulative spectral energy density,
\begin{equation}
\frac{d \widehat{E}_k}{dt} \approx \frac{d}{dt}\left(\frac{|\widehat{u}_k|^2}{2} + \frac{|\widehat{p}_k|^2}{2}\right) + \Re\left(\widehat{p}_{-k}\frac{d\widehat{g}_k}{dt}\right),
\label{eq: SpectralEnergyRate}
\end{equation}
and $\widehat{g}$ is the Fourier transform of $g(\pFluct)$ given by,
\begin{equation}
g(\pFluct) = \frac{\gamma}{\gamma-1}\left(\left(1+\gamma \pFluct\right)^{1/\gamma} - 1 -\pFluct \right).
\end{equation}
The spectral energy $\widehat{E}_k$ is given by, 
\begin{equation}
\widehat{E}_k = \frac{|\widehat{u}_k|^2}{2} + \frac{|\widehat{p}_k|^2}{2} + \Re \left(\widehat{p}_{-k}\left(\widehat{\frac{f(\pFluct)}{\pFluct}}\right)_k\right).
\label{eq: spectral_energy}
\end{equation}
It is noteworthy that the correction in spectral energy does not follow directly from the nonlinear correction function $f(\pFluct)$ derived in the physical space. In Eq.~\eqref{eq: SpectralEnergyRate}, we have made the following approximation,
\begin{equation}
\frac{d}{dt}\left( \Re \left(\widehat{p}_{-k}\left(\widehat{\frac{f(\pFluct)}{\pFluct}}\right)_k\right)\right) \approx \Re\left(\widehat{p}_{-k}\frac{d\widehat{g}_k}{dt}\right)
\end{equation}
Second term $\widehat{\Pi}_k$ in Eq.~\eqref{eq: spectral_conservation} is the flux of spectral energy density from wavenumbers $|k'|\leq k$ to $|k'| > k$ and is given by, 
\begin{align}
\widehat{\Pi}_k &= \sum_{|k'|\leq k}\Re\Big(\widehat{p}_{-k'}\left(\frac{\partial (\widehat{\uFluct g})}{\partial x}\right)_{k'} +  \widehat{p}_{-k'}\left(\widehat{\uFluct\frac{\partial \pFluct}{\partial x}}\right)_{k'} +\nonumber \\
&\frac{1}{2}\widehat{u}_{-k'}\reallywidehat{\frac{\partial}{\partial x}\left(\uFluct ^2 - \pFluct ^2\right)_{k'}}\Big).
\label{eq: SpectralFlux}
\end{align}
Finally, the spectral dissipation $\widehat{\mathcal{D}}_k$ is given by, 
{\color{black}{
\begin{align}
\widehat{\mathcal{D}}_k = \nu_0\frac{\gamma - 1}{Pr}&\Re\left(\widehat{p}_{-k}\left(\reallywidehat{\left(1 + \frac{\partial g}{\partial \pFluct}\right)\left(\frac{\partial^2 \pFluct}{\partial x^2}\right)}\right)_{k}\right) \nonumber \\
&- {\frac{16\pi^2}{3}\nu_0} k^2|\widehat{u}_k|^2.
\label{eq: SpectralDissipation}
\end{align}
}}
Detailed derivation of Eqs.~\eqref{eq: spectral_conservation}-\eqref{eq: SpectralDissipation} is given in appendix~\ref{sec: appedixB}. Figure~\ref{fig: LengthScalesSummary} summarizes the typical shape of the spectral energy $\widehat{E}_k$, spectral energy flux $\widehat{\Pi}_k$, and the spectral dissipation $\widehat{\mathcal{D}}_k$ along with the relative positions of the three relevant length scales, the integral length scale, $\ell$, the Taylor microscale, $\lambda$, and the Kolomogorov length scale $\eta$ in the spectral space. The spectro-temporal evolution of any configuration of nonlinear acoustic waves can be quantified utilizing these length scales and the respective evolution in time which is discussed in detail in the subsections below. Table~\ref{tab: LengthScalesTable} summarizes these length scales and the characteristic spectral range. {In further sections, we discuss all the spectral quantities as functions of absolute value of wavenumbers and drop the $|.|$ notation for convenience.}

The spectral energy flux $\widehat{\Pi}_k$, defined in Eq.~\eqref{eq: SpectralFlux}, is in terms of interactions of the Fourier coefficients of the pressure $\widehat{p}_k$ and velocity $\widehat{u}_k$ perturbations. For compact support or periodic perturbations, $\widehat{\Pi}_k$ approaches zero in the limit of very large wavenumbers $k\rightarrow \infty$, 
\begin{equation}
\lim_{k\rightarrow\infty} \widehat{\Pi}_{k} = \left\langle \frac{\partial I}{\partial x}\right\rangle = 0.
\end{equation} 
The last two terms in Eq.~\eqref{eq: SpectralFlux} result in $\widehat{\Pi}_{k}\rightarrow 0$ for large $k$ for general acoustic phasing. Hence, they are most relevant in SW and AWT cases. In a pure traveling wave (TW), $\uFluct = \pFluct$ at first order due to which the last two terms in Eq.~\eqref{eq: SpectralFlux} become negligible. Furthermore, the sequence of $\widehat{\Pi}_k$ also converges monotonically i.e.,
\begin{equation}
 \lim_{k\rightarrow\infty}\left(\widehat{\Pi}_{k-1} - \widehat{\Pi}_{k}\right) \rightarrow 0^{+},
 \label{eq: MonotonicConv}
\end{equation}
as shown in Figs.~\ref{fig: LengthScalesSummary}$(c)$ and~\ref{fig: EnergyFLux}. The flattening of the spectral energy flux $\widehat{\Pi}_k$ (Eq.~\eqref{eq: MonotonicConv}) begins at a specific wavenumber $k_\delta$ associated to the {Kolmogorov length scale}, $\eta$, as shown in Fig.~\ref{fig: LengthScalesSummary}. The spectral energy $\widehat{E}_k$ deviates off the $k^{-2}$ decay near the wavenumber $k_\delta$. Figure~\ref{fig: EnergyFLux} shows the spectro-temporal evolution of the spectral energy $\widehat{E}_k$ and the flux $\widehat{\Pi}_k$ for TW, SW, and AWT prior to formation of shock waves.

For TW, $\widehat{\Pi}_k$ increases in time due to spectral broadening. In SW, $\widehat{\Pi}_k$, while increasing, also oscillates at low wavenumbers due to the periodic collisions of oppositely propagating shocks. A combination of these processes takes place in a randomly initialized smooth finite amplitude perturbation, which at later times develops into AWT. At later times, nonlinear waves in all three configurations fully develop in to shock waves. Up to the shock formation, the spectral dynamics of all configurations simply involve increase of the spectral flux $\widehat{\Pi}_k$. The dimensionless shock formation time $\tau$ can be estimated as,
\begin{equation}
\tau = \frac{2}{(\gamma - 1)A_{\mathrm{rms},0}}.
\label{eq: SteepeningTime}
\end{equation}
Upon shock formation, the dynamic evolution of TW and SW remains phenomenologically identical. The isolated shocks propagate and the total perturbation energy of the system decays due to thermoviscous dissipation localized around the shock wave. {\color{black}{However, for AWT, along with collisions of oppositely propagating shocks, those propagating in the same direction coalesce due to differential propagating speeds. As we discuss below, this modifies the energy decay and spectral energy dynamics in AWT significantly compared to  TW and SW.}}

In the sub-sections below, we elucidate the energy dynamics before and after shock formation for TW, SW, and AWT. To this end, we define and discuss the relevant length scales as mentioned above, namely: the Taylor microscale $\lambda$, the integral length scale $\ell$, and the Kolmogorov length scale $\eta$. Particular focus is given to the AWT case due to modified dynamics caused by shock coalescence.

\begin{figure*}[!t]
\centering
\includegraphics[width=\textwidth]{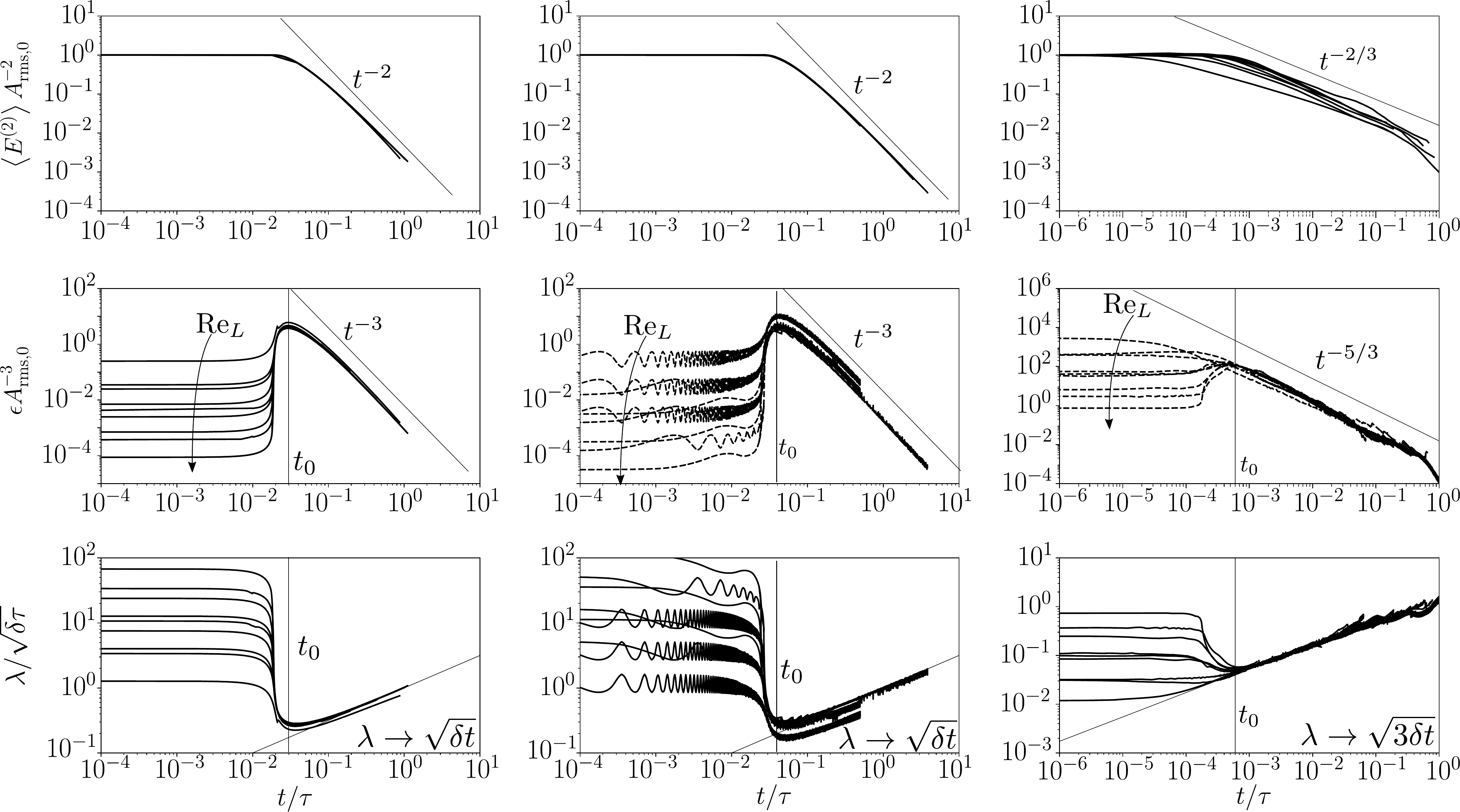}
\put(-510,290){$(a)$}
\put(-330,290){$(b)$}
\put(-160,290){$(c)$}
\caption{Temporal evolution of scaled total energy <$E^{(2)}$> ${A^{-2}_{\mathrm{rms},0}}$ (--) (top), dissipation rate $\epsilon {A^{-3}_{\mathrm{rms},0}}$ ($--$) (mid) and normalized Taylor microscale $\lambda/\sqrt{\delta \tau}$ (bottom) for TW ($a$), SW ($b$) and AWT ($c$) against the scaled time $t/\tau$ for varying perturbation Reynolds number $\mathrm{Re}_L$. The time $t_0$ signifies fully broadened spectrum of the perturbation field.}
\label{fig: EnergyDecayAll}
\end{figure*}

\subsection{Taylor microscale}
In hydrodynamic turbulence, the Taylor microscale $\lambda$ separates the inviscid length scales from the viscous length scales~\cite{tennekes1972first,pope2000turbulent}. Due to the spectral energy cascade in planar nonlinear acoustics, we note that the spectral energy varies as $\widehat{E}_k\sim k^{-2}$ due to the formation of shocks and the spectral dissipation due to thermoviscous diffusion varies as $\widehat{\mathcal{D}}_k\sim k^2 \widehat{E}_k$. Consequently, the dissipation acts over most of the length scales with $k>k_E$ (Fig.~\ref{fig: LengthScalesSummary}$(d)$), unlike hydrodynamic turbulence where the viscous dissipation dominates only the smaller length scales~\cite{tennekes1972first, pope2000turbulent}. As shown in Fig.~\ref{fig: LengthScalesSummary}$(c)$, length scales in the range $(k_E, k_{\delta})$ exhibit both dissipation $\widehat{\mathcal{D}}_k$ and energy transfer $\widehat{\Pi}_k$. For $k>k_{\delta}$, $\widehat{\Pi}_k$ begins to converge monotonically to 0 and the interval $(k_{\delta}, 1/\eta)$ primarily exhibits dissipation $\widehat{\mathcal{D}}_k$ only. The Taylor microscale $\lambda$ quantifies the length scale associated to the whole dissipation range. 

Utilizing the definition of the total perturbation energy $\left\langle E^{(2)} \right\rangle$ and the dissipation rate $\epsilon$ (cf. Eq.~\eqref{eq: LyapunovFunction}), the microscale $\lambda$ can be defined as, 
\begin{equation}
\lambda(t) = \sqrt{\frac{2\delta\left\langle E^{(2)}\right\rangle }{\epsilon}},
\label{eq: Microscale}
\end{equation}
where $\delta$ is the thermoviscous diffusivity, given by, 
{\color{black}{
\begin{equation}
\delta = \nu_0\left(\frac{4}{3} + \frac{\gamma - 1}{\mathrm{Pr}}\right).
\label{eq: EffectiveDiff}
\end{equation}
}}
Equation~\eqref{eq: Microscale} indicates that the Taylor microscale can be identified as the geometrical centroid of full energy spectrum, i.e.
\begin{equation}
\lambda \sim \sqrt{\frac{\sum_k \widehat{E}_k}{\sum_k k^2 \widehat{E}_k}}.
\label{eq: lengthscales_centroids}
\end{equation}

As the smaller length scales (higher harmonics) are generated, the dissipation rate $\epsilon$ tends to increase reaching a maximum in time. The increase of dissipation rate $\epsilon$ implies decrease of the length scale $\lambda$ in time. Minima of $\lambda$ indicates the fully-broadened spectrum of energy limited by the thermoviscous diffusivity at very large scales. Further spatio-temporal evolution of the system is dominated by dissipation thus indicating the purely diffusive nature of the Taylor microscale, i.e.,
\begin{equation}
 \lambda\rightarrow C\sqrt{\delta t}.
\label{eq: TaylorMicroscaleAssy}
\end{equation}
The temporal evolution of $\lambda$ is qualitatively similar for TW, SW, and AWT, the constant $C$ in Eq.~\eqref{eq: TaylorMicroscaleAssy} differs for TW and SW compared with AWT due to the different spatial structure of perturbations. The time $t_0$ at which $\lambda$ reaches minimum signifies {fully developed nonlinear acoustic waves}. In case of AWT, it signifies fully developed acoustic wave turbulence. 

Figure~\ref{fig: EnergyDecayAll} shows the decay of scaled total perturbation energy $\left\langle E^{(2)}\right\rangle A_{\mathrm{rms},0}^{-2}$ and total dissipation rate $\epsilon A_{\mathrm{rms},0}^{-3}$ for the TW, SW, and AWT. We note that the total energy decays as a power law $t^{-2}$ for both TW and SW, whereas, for AWT, the initial decay law is $t^{-2/3}$. Asymptotic evolution (at large $t$) of the Taylor microscale follows from the decay laws as $\lambda = \sqrt{\delta t}$ and $\lambda = \sqrt{3\delta t}$ respectively. Since energy decay law of a single harmonic traveling and standing waves is rather trivial, we focus primarily on the AWT case for further discussion.

\subsection{Integral length scale}
\begin{figure}[!b]
\centering
\includegraphics[width=\linewidth]{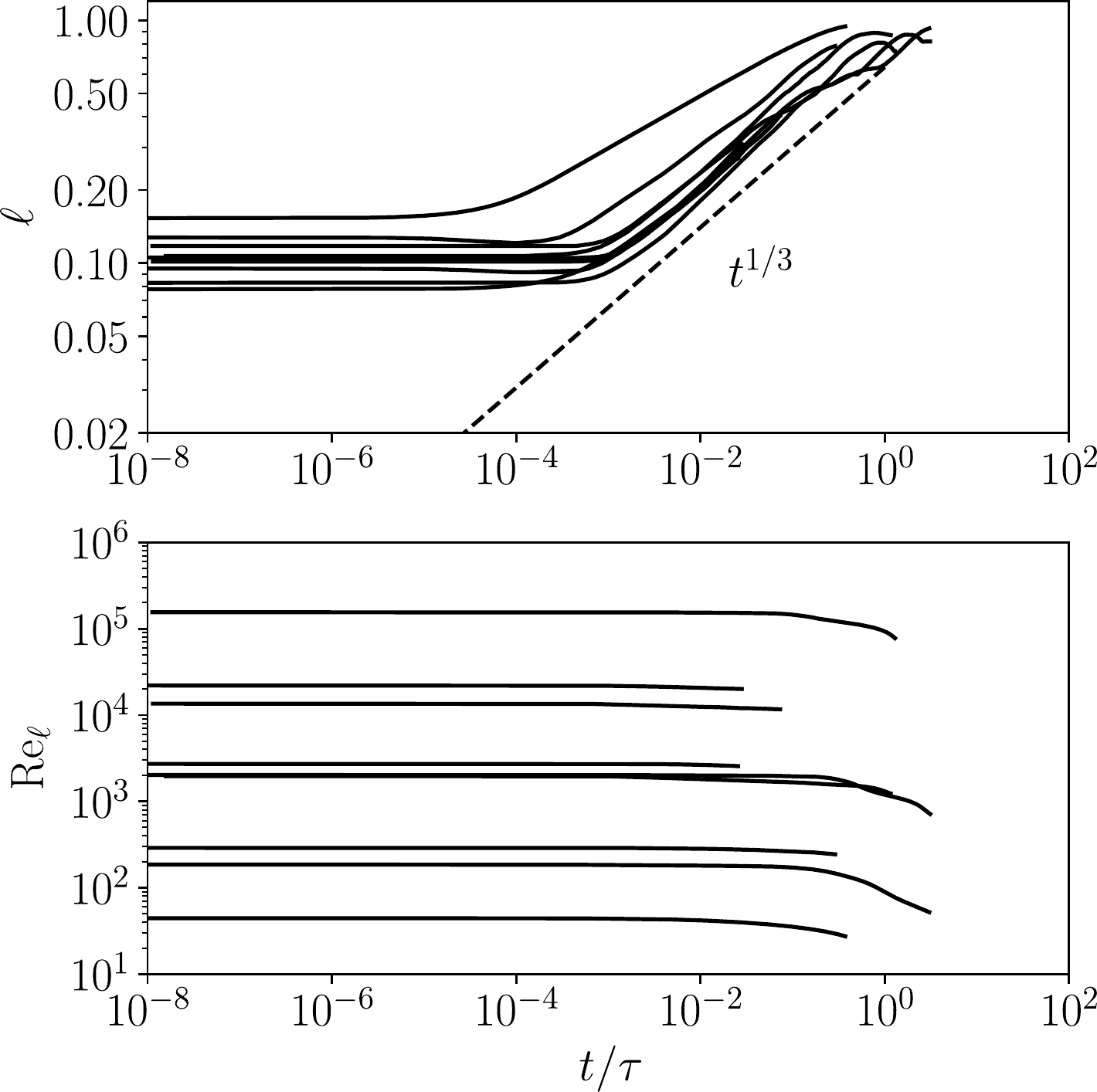}
\put(-250,250){$(a)$}
\put(-250,125){$(b)$}
\caption{Evolution of the integral length scale $\ell$ $(a)$ and the Reynolds number $\mathrm{Re}_{\ell}$ ($b$) defined in Eqs.~\eqref{eq: IntegralLengthScale} and~\eqref{eq: IntegralReynoldsNumber}, respectively, for all the cases of AWT considered. For small thermoviscous diffusivity, $\ell$ increases approximately as $t^{1/3}$ before saturating to the dimensionless domain length $L=1$ and $\mathrm{Re}_{\ell}$ remains approximately constant.}
\label{fig: IntegralLengthScaleEvolution}
\end{figure}
We identify the integral length scale $\ell$ as the characteristic length scale of the energy containing scales. In general, random {smooth} broadband noise (AWT) develops into an ensemble of shocks, propagating left and right in a one-dimensional system. For an ensemble of shock waves distributed spatially along a line, $\ell$ corresponds to the characteristic distance between consecutive shock waves traveling in the same direction, as shown schematically in Fig.~\ref{fig: LengthScalesSummary}.
Formally, we define $\ell$ as, 
\begin{equation}
\ell = \sqrt{\frac{\sum_k \frac{\widehat{E}_k}{k^2}}{\sum_k \widehat{E}_k}},
\label{eq: IntegralLengthScale}
\end{equation}
which is identical to the integral length scale defined in Burgers turbulence~\cite{Gurbatov_JFM_1997}. {\color{black}{Definition in Eq.~\eqref{eq: IntegralLengthScale} yields the centroid wavenumber of the initial energy spectrum (unlike Taylor microscale, which corresponds to the full energy spectrum) and hence is characteristic of the large length scales of fully developed AWT.}}
To elucidate the evolution of the total perturbation energy $\left\langle E^{(2)}\right\rangle$ utilizing the integral length scale, we assume the following model spectral energy density $\widehat{E}_k$,
\begin{equation}
\widehat{E}_k = \begin{cases} 
      C_1 k^n & k_0\leq k \leq k_E \\
     C_2 k^{-2} & k_{\delta}> k > k_E
   \end{cases},
\label{eq: SpectralEnergyAssump}
\end{equation}
where $k^n$ corresponds to the shape of initialized energy spectrum in the range $(k_0,k_E)$ (Fig.~\ref{fig: LengthScalesSummary}). In this work, we only utilize the white noise initialized AWT cases which correspond to $n=0$ (see Table~\ref{tab: InitialSpectra}). Moreover,
\begin{equation}
C_2 = C_1 k_E^{n+2}.
\end{equation}
The wave numbers $k_E$ and $k_\delta$ vary in time due to decaying energy. By definition, the mean of perturbations is zero. Hence, the smallest wavenumber containing energy $k_0$ (cf. Fig.~\ref{fig: LengthScalesSummary}) is the reciprocal of the domain length $L$, i.e., 
\begin{equation}
k_0 = 1/L.
\end{equation}

We note that the above model spectral energy $\widehat{E}_k$ holds for two primary reasons. Firstly, the energy cascade results in the $k^{-2}$ decay of the spectral energy $\widehat{E}_k$ due to formation of shock waves~\cite{Gurbatov_JFM_1997}. In the limit of vanishing viscosity $\delta \rightarrow 0 $, such decay extends up to $k\rightarrow \infty$ in which case the developed shock waves render the system $C_0$ discontinuous. Secondly, the shape of the spectral energy $\widehat{E}_k$ for $k\rightarrow k_0$ corresponds to $k^{n}$, which is also the shape of initial energy spectral at time $t=0$. Such argument corresponds to the concept of \emph{permanence of large eddies} in hydrodynamic turbulence~\cite{Batchelor1953theory}, which in spectral space can be written as, 
\begin{equation}
\widehat{E}_k(t) \approx \widehat{E}_k(t=0),~~\mathrm{as}~~k\rightarrow k_0.
\label{eq: PLE}
\end{equation}
Gurbatov \emph{et al.}~\cite{Gurbatov_JFM_1997} utilized a similar argument in the context of Burgers turbulence. Combining the Eqs.~\eqref{eq: IntegralLengthScale}-\eqref{eq: PLE}, the integral length scale $\ell$ is given by, 
\begin{align}
\ell &\approx 
\begin{cases} 
     \sqrt{\frac{n+1}{n-1}\left(\frac{k^{n-2}_E + k^{n-3}_Ek_0 + \cdots k^{n-2}_0}{k^{n}_E + k^{n-1}_Ek_0 + \cdots k^{n}_0}\right)} & n\neq 1 \\
     \sqrt{\frac{2\ln(k_E/k_0)}{k^2_E - k^2_0}} & n = 1.
   \end{cases},
\label{eq: IntegralLengthN}
\end{align}
where we have used the simplifying approximation of $k_\delta \gg k_E$. We note that the Eq.~\eqref{eq: IntegralLengthN} indicates the dependence of $\ell$ and consequently the energy decay law on $n$. In the present work, we perform numerical simulations for an uncorrelated white noise (filtered) which corresponds to $n=0$, and 
\begin{equation}
\ell \approx \frac{1}{\sqrt{k_0 k_E}}.
\label{eq: IntegralApprox}
\end{equation}
As a result of \emph{permanence of large eddies}, the decay of energy in the initial regime of AWT is associated only to the decreasing $k_E$ or increasing integral length scale $\ell$. Integrating Eq.~\eqref{eq: SpectralEnergyAssump} in the spectral space and differentiating in time yields (for $n=0$ in the current simulations), 
\begin{align}
\frac{d \left\langle E^{(2)} \right\rangle}{dt} &= C_1\left(2\frac{dk_E}{dt}\left(1 - \frac{k_E}{k_\delta}\right) + \left(\frac{k_E}{k_{\delta}}\right)^2\frac{dk_\delta}{dt}\right)  \\ &\approx -\frac{2C_1}{k_0 \ell^3}\frac{d\ell}{dt}.
\label{eq: EdotLdot}
\end{align}
Above relation shows that derivation of the energy decay power law amounts to finding the kinetic equations of the integral length scale $\ell$ and the limiting wavenumber $k_\delta$. For TW and SW, $\ell$ remains constant by definition. Consequently, the energy decay rate only depends on decrease of wavenumber $k_\delta$ and the coefficient $C_2$ due to the thermoviscous diffusion (cf. Eq.~\eqref{eq: dEdt1}
). However, for an ensemble of shock waves in AWT, $\ell$ increases monotonically in time, as shown in Fig.~\ref{fig: IntegralLengthScaleEvolution}$(a)$ due to the coalescence of shock waves propagating in the same direction. At large times, the domain consists of only two shock waves propagating in opposite directions. 

In the context of Burgers turbulence in an infinite one-dimensional domain, Burgers~\cite{Burgers1974nonlinear} and Kida~\citep{Kida_1979_JFM} have derived the appropriate asymptotic evolution laws for the integral length scale $\ell$ based on the dimensional arguments. However, in the present work, the finiteness of the domain renders the asymptotic analysis infeasible. Our numerical results indicate that $\ell\sim t^{1/3}$ ($k_E\sim t^{-2/3}$) for randomly distributed shock waves at various $\mathrm{Re}_L$ values considered, as shown in Fig.~\ref{fig: IntegralLengthScaleEvolution}$(a)$. Equations~\eqref{eq: EdotLdot} and~\eqref{eq: IntegralApprox} show that such scaling is consistent with the observed energy decay law $\left\langle E^{(2)} \right\rangle \sim t^{-2/3}$ thus validating the result in Eq.~\eqref{eq: EdotLdot}. It is noteworthy that decay $k_E\sim t^{-2/3}$ is a result analogous to the one discussed in Burgers turbulence~\cite{Burgers1974nonlinear,Kida_1979_JFM,Gurbatov_JFM_1997} considered in an infinite one-dimensional domain. {\color{black}{Due to infinitely long domain, the average distance between the shocks approaches $1/k_E$ (not $\ell$) simply due to larger number of shocks in the domain separated by the distance $1/k_E$ since $k_E$ corresponds to the largest wavenumber carrying initial energy, thus implying that mean distance between the shocks increases as $t^{2/3}$ as noted by Burgers~\cite{Burgers1974nonlinear}.}}

Based on the integral length scale, the Reynolds number $\mathrm{Re}_{\ell}$ can be defined as, 
\begin{equation}
\mathrm{Re}_{\ell} = \mathrm{Re}_{L}\ell,
\label{eq: IntegralReynoldsNumber}
\end{equation}
which captures the ratio of the diffusive time scale to the wave turbulence time. Upon formation of shock waves, the perturbation energy decays due to coalescence. Shock waves coalesce locally thus increasing the characteristic separation between the shock waves thus causing $\ell$ to increase. In this regime, the Reynolds number $\mathrm{Re}_{\ell}$ remains constant (Fig.~\ref{fig: IntegralLengthScaleEvolution}$(b)$) which denotes that the ratio of shock coalescence time scale $(\ell L^*)/(a^*_0A_{\mathrm{rms}})$ and the diffusive time scale $(\ell L^*)^2/\nu^*_0$ remains constant. As the wave turbulence decays further, $\ell \rightarrow L$ with continued decay of energy. Consequently, $\mathrm{Re}_{\ell}$ also begins to decay.

\subsection{Kolmogorov length scale}
For spectral energy $\widehat{E}_k\sim k^{-2}$ over the intermediate range of wavenumbers, $k\in (k_E, k_\delta)$
(cf. Fig.~\ref{fig: LengthScalesSummary}), the Taylor microscale can be estimated as, 
\begin{equation}
\lambda\sim \frac{1}{\sqrt{k_E k_\delta}},
\label{eq: TaylorMicroscaleApp}
\end{equation}
utilizing the Eq.~\eqref{eq: lengthscales_centroids}. Equation~\eqref{eq: TaylorMicroscaleApp} shows that $\lambda$, despite being a dissipative scale, is not the smallest scale generated due to the energy cascade.
Analogous to the hydrodynamic turbulence, we define the Kolmogorov length scale $\eta$~\cite{tennekes1972first} as the smallest length scale generated as a result of the acoustic energy cascade. The length scale $\eta$ can be approximated by the balance of nonlinear steepening and energy dissipation, i.e.,
\begin{equation}
\frac{{A}^2_{\mathrm{rms}}}{\eta} \sim \delta\frac{A_{\mathrm{rms}}}{\eta^2}, \quad \eta \sim \frac{\delta}{{A}_{\mathrm{rms}}},
\label{eq: KolomogorovScale}
\end{equation}
where $A_{\mathrm{rms}}$ is defined in Eq.~\eqref{eq: PcDef}. Figure~\ref{fig: LengthScalesSummary} illustrates the integral length scale $\ell$ and the Kolmogorov length scale $\eta$ in a typical AWT field. Visual inspection indicates $\ell \gg \eta$ which is as expected. We note that $\eta$ and $1/k_{\delta}$ evolve in time similarly, differing only by a constant value. For AWT, this is immediately realizable since, Eq.~\eqref{eq: TaylorMicroscaleApp} shows that $k_\delta \sim t^{-1/3}$ and Eq.~\eqref{eq: KolomogorovScale} shows that $\eta \sim t^{1/3}$ which implies $k_\delta \eta$ remains constant when the energy decays. 
\begin{figure}
\centering
\includegraphics[width=\linewidth]{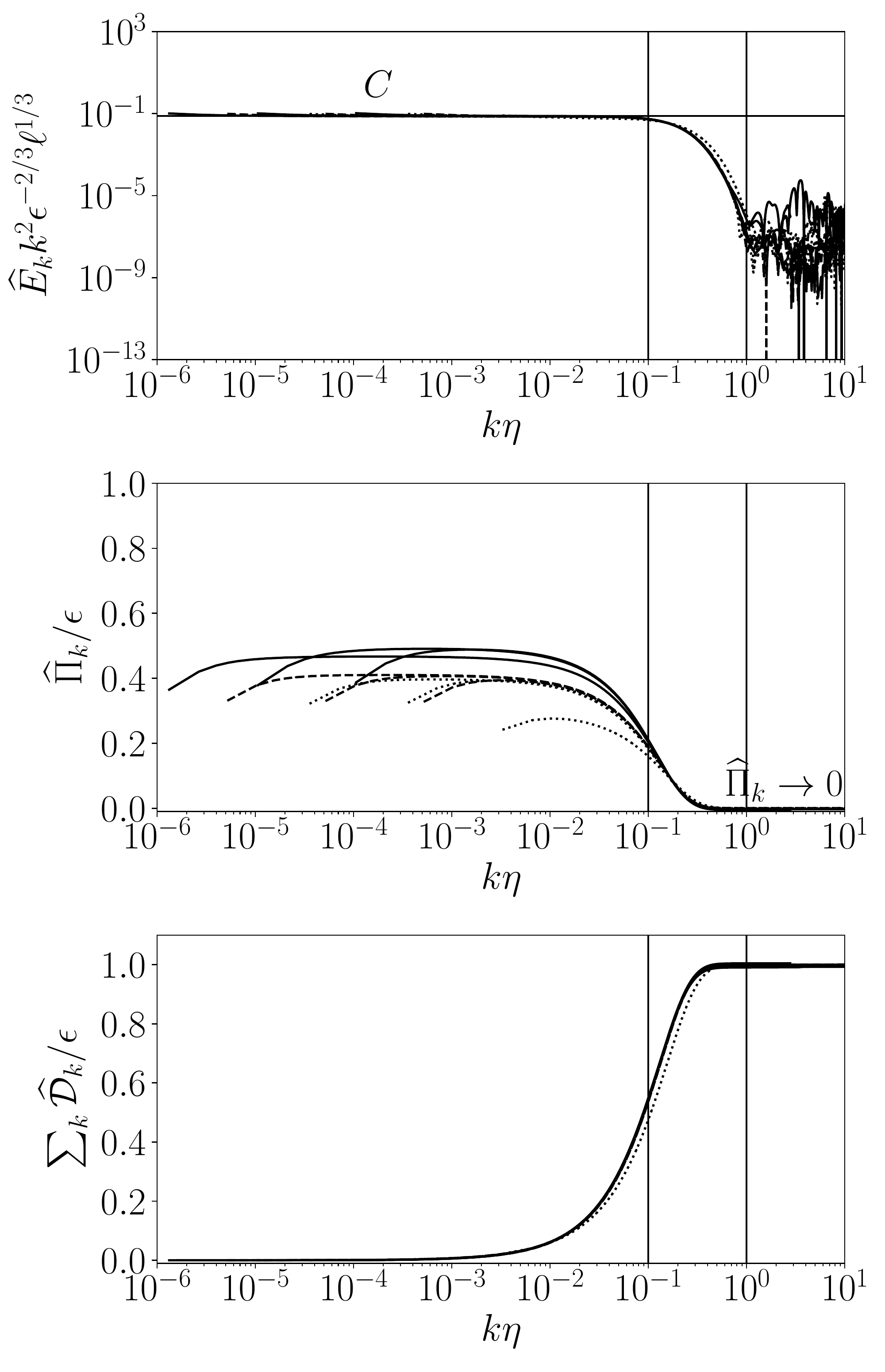}
\put(-240,370){$(a)$}
\put(-240,250){$(b)$}
\put(-240,120){$(c)$}
\caption{{\color{black}{Fully developed spectra of compensated energy $(a)$, spectral energy flux $(b)$, and cumulative dissipation $(c)$ for TW at time instant $t_0\approx0.03$. Harmonics with wavenumbers such that $k\eta < 1$ contain all the energy. The spectral energy flux vanishes at $k\eta \approx 1$ thus indicating numerical resolution of all the energy containing harmonics. The marked regime $0.1<k\eta<1$ signifies the dissipation range. The constant $C\approx 0.075$. (--) $A_{\mathrm{rms},0}=10^{-1}$; ($--$) $A_{\mathrm{rms},0}=10^{-2}$; ($\cdots$) $A_{\mathrm{rms},0}=10^{-3}$}}}
\label{fig: SpectralFlux_Traveling}
\end{figure}
For TW and SW, the spectral energy given by Eq.~\eqref{eq: SpectralEnergyAssump} corresponds to the degenerate case of $k_0 = k_E = 1$. For such a form of spectral energy, the energy evolution (cf. Eq.~\eqref{eq: EdotLdot}) changes to, 
\begin{equation}
 \frac{d\left\langle E^{(2)}\right\rangle}{dt} = \frac{1}{k_0}\frac{dC_2}{dt}\left(1 - \frac{k_0}{k_\delta}\right) + \frac{C_2}{k^2_\delta}\frac{dk_\delta}{dt}.
 \label{eq: dEdt1}
\end{equation}
As shown in Fig.~\ref{fig: EnergyDecayAll}, the Taylor microscale $\lambda \rightarrow \sqrt{\delta t}$. Consequently, for $k_E = k_0$ constant, Eq.~\eqref{eq: TaylorMicroscaleApp} shows that $k_\delta \sim t^{-1}$. Equation~\eqref{eq: dEdt1} shows that the decay of perturbation energy is due to decay in $C_2$ and $k_{\delta}$. Our numerical results (cf.~Fig.~\ref{fig: EnergyDecayAll}) show that for TW and SW, $\left\langle E^{(2)} \right\rangle \sim t^{-2}$ which suggests that $C_2\sim t^{-2}$ for $k_\delta\gg 1$ from Eq.~\eqref{eq: dEdt1}. Hence, the compensated energy spectrum $k^2\widehat{E}_k\sim t^{-2}$ for both TW and SW indicating that dissipation $\widehat{\mathcal{D}}_k$ remains active over all the length scales $k>k_0$ while the energy decays. 

Equation~\eqref{eq: KolomogorovScale} shows that the Reynolds number based on the Kolmogorov length scale or the shock thickness $Re_{\eta} = \eta \mathrm{Re}_L$ remains constant in time, 
{\color{black}{
\begin{equation}
 Re_{\eta} = \frac{\rho^*_0 a^*_0 L^* \eta A_{\mathrm{rms}}}{\mu^*_0} = \frac{4}{3} + \frac{\gamma - 1}{Pr}.
\end{equation}
}}
Above relation shows that $Re_{\eta} = \mathcal{O}\left(1\right)$ indicating that $\eta$ is the length scale at which diffusion dominates the nonlinear wave steepening.\section{Scaling of spectral quantities}
\label{sec: SpectralScaling}

In this section, we discuss the variation and scaling of the energy $\widehat{E}_k$, the spectral energy flux $\widehat{\Pi}_k$, and the cumulative dissipation $\sum_{k'<k} \widehat{\mathcal{D}}_{k'}$ for high amplitude TW, SW, and AWT cases utilizing the length scale analysis presented in the previous sections. We show that the spectral energy $\widehat{E}_k$ and the cumulative dissipation $\sum_{k'<k} \widehat{\mathcal{D}}_{k'}$ for all the cases can be collapsed on to a common structure versus the reduced wavenumber $k\eta$ however, the flux $\widehat{\Pi}_k$ lacks such a universality. 

 As discussed in previous section (cf. Eq.~\eqref{eq: dEdt1}), the decay of total energy $\left\langle E^{(2)}\right\rangle$ and dissipation rate $\epsilon$ for TW is given by, 
\begin{equation}
\left\langle E^{(2)}\right\rangle \sim t^{-2},~\mathrm{and}~\epsilon\sim t^{-3},
\label{eq: EnergyTimeTraveling}
\end{equation}
which are well known results for the Burgers equation as well~\citep{Bec_PhyRep_2007}.
\begin{figure}[!b]
\centering
\includegraphics[width=\linewidth]{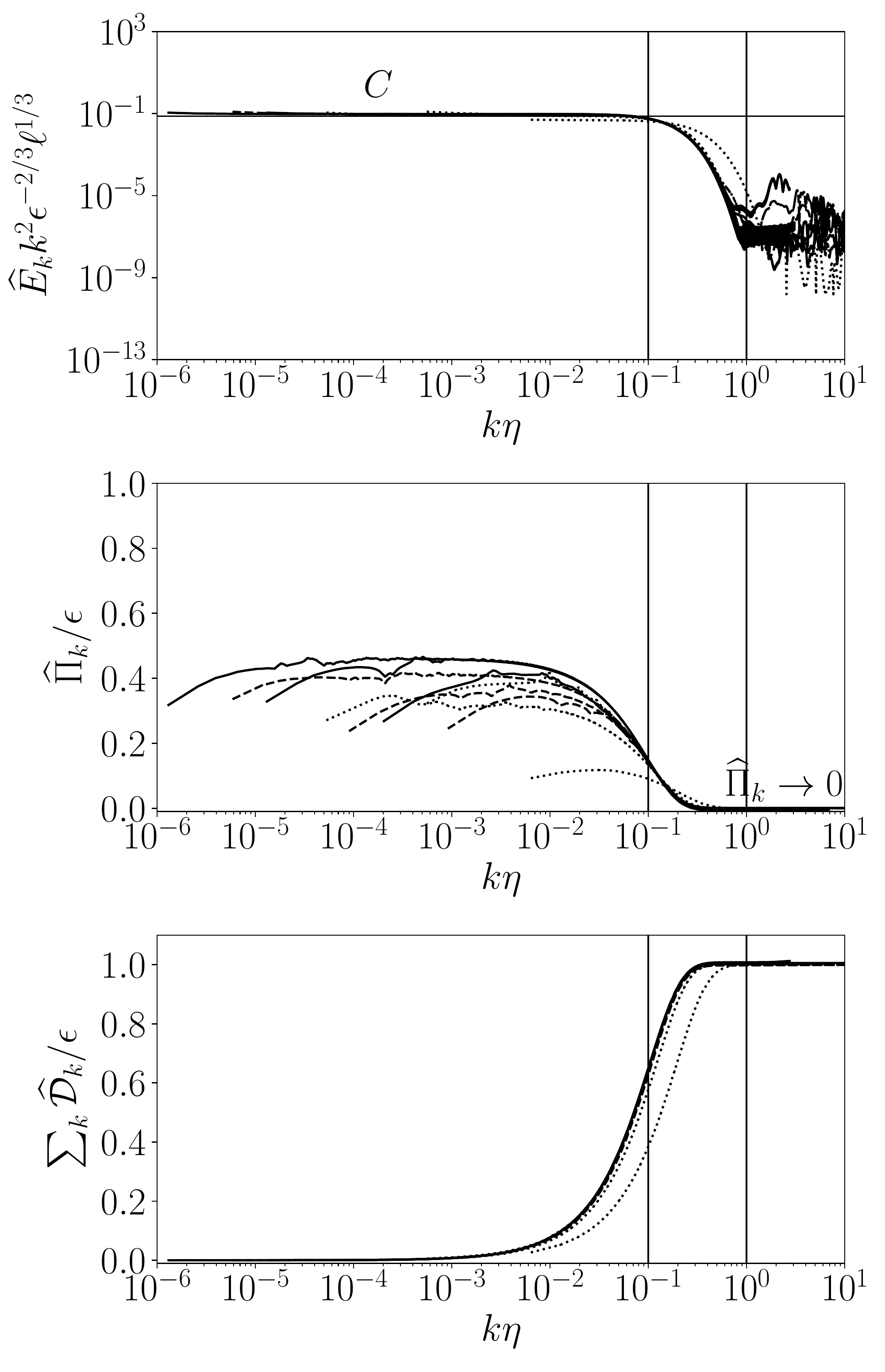}
\put(-240,370){$(a)$}
\put(-240,250){$(b)$}
\put(-240,120){$(c)$}
\caption{Fully developed spectra of compensated energy $(a)$, spectral energy flux $(b)$, and cumulative dissipation $(c)$ for SW averaged over one time cycle after $t_0\approx0.04$. Harmonics with wavenumbers $k\eta < 1$ contain all the energy. The spectral energy flux vanishes at $k\eta \approx 1$ thus indicating numerical resolution of all the energy containing harmonics. The marked regime $0.1<k\eta<1$ signifies the dissipation range. The constant $C\approx 0.075$. (--) $A_{\mathrm{rms},0}=10^{-1}$; ($--$) $A_{\mathrm{rms},0}=10^{-2}$; ($\cdots$) $A_{\mathrm{rms},0}=10^{-3}$}
\label{fig: SpectralFlux_Standing}
\end{figure}

While the results in Eq.~\eqref{eq: EnergyTimeTraveling} are well known, we note that such power law decay results in a universally constant structure of shock waves in the spectral space, as shown in Figs.~\ref{fig: SpectralFlux_Traveling} and~\ref{fig: SpectralFlux_Standing}. Utilizing the estimate of Kolmogorov length scale $\eta$ given in Eq.~\eqref{eq: KolomogorovScale}, the energy dissipation rate $\epsilon$ and the Kolmogorov length scale $\eta$ can be related as,
\begin{equation}
\epsilon \sim \frac{{A_{\mathrm{rms}}^3}}{\ell},~\mathrm{and}~ \eta \sim \frac{\delta}{\left(\epsilon \ell\right)^{1/3}}.
\end{equation}
{\color{black}{Hence, the energy spectrum $\widehat{E}_k$ can be written in the following collapsed form (Fig.~\ref{fig: SpectralFlux_Traveling}$a$).
\begin{equation}
\widehat{E}_kk^{2}\epsilon^{-2/3}\ell^{1/3} \sim C F(k\eta).
\label{eq: SpectralForm}
\end{equation}
In Eq.~\eqref{eq: SpectralForm}, the integral length scale $\ell$ is used for making the left hand expression dimensionless}}. For TW and SW, the integral length scale $\ell$ remains constant by definition ($\ell = L$). Hence, $C$ in Eq.~\eqref{eq: SpectralForm} is constant and can be attributed to the Kolmogorov's universal equilibrium theory for hydrodynamic turbulence. $F(.)$ is a function which decays as the reduced wavenumber $k\eta$ increases to 1. From the numerical simulations for cases listed in Table~\ref{tab: test_cases} we obtain,
\begin{equation}
C \approx 0.075.
\label{eq: ConstantValue}
\end{equation} 
Scaling of $\widehat{\Pi}_k$ with the energy dissipation rate $\epsilon$ shows the relative magnitude of spectral energy flux compared to the energy dissipation.{\color{black}{ For increasing Reynolds numbers $\mathrm{Re}_L$, we note that $\widehat{\Pi}_k/\epsilon$ increases but still remains less than 1 in the energy transfer and dissipation range, as shown in Fig.~\ref{fig: SpectralFlux_Traveling}$(b)$. This highlights the primary difference between energy spectra of nonlinear acoustic waves and hydrodynamic turbulence, in which, the energy transfer range does not exhibit viscous dissipation~\cite{pope2000turbulent}. However, in nonlinear acoustics, the dissipation occurs over all the smaller length scales which do not contain energy initially (Fig.~\ref{fig: LengthScalesSummary}$(d)$). Moreover, for $k\eta\approx 0.1$, the flux $\widehat{\Pi}_k$ rapidly approaches to zero. In the regime $k\eta>0.1$, scaled cumulative dissipation $\sum_{k'<k} \widehat{\mathcal{D}}_{k'}/\epsilon\rightarrow 1$ as $k\eta\rightarrow 1$.}}

Such functional forms of spectral energy, spectral energy flux, and cumulative dissipation can also be realized for the SW case. At later times, the nonlinear evolution results in two opposite traveling shock waves which collide with each other twice in one time period. Such collisions cause instantaneous peaks in the dissipation rate $\epsilon$ and corresponding oscillations in the Taylor microscale $\lambda$, as shown in Fig.~\ref{fig: EnergyDecayAll}. However, the total energy $\left\langle E^{(2)} \right\rangle$ decays monotonically by definition. In the spectral space, such collisions generate periodic oscillations in the spectral energy flux $\widehat{\Pi}_k$, as shown in Fig.~\ref{fig: EnergyFLux}. Averaging over one such time cycle yields the energy spectra forms similar to that for TW, as shown in Fig.~\ref{fig: SpectralFlux_Standing}. Such cycle averaging is allowed since the total energy $\left\langle E^{(2)}\right\rangle$ and the dissipation rate $\epsilon$ decay such that averaged behavior is identical to the one of traveling waves. Furthermore, the value of the constant $C$ is identical for SW.{\color{black}{ We further note that for the case with lowest Reynolds number $\mathrm{Re}_L$ ($\nu_0 = 1.836\times10^{-5}$ and $A_{0,\mathrm{rms}} = 10^{-3}$), the spectra exhibit energy for $k\eta > 1$ (Fig.~\ref{fig: SpectralFlux_Standing}$(a)$ since the Eq.~\eqref{eq: KolomogorovScale} underpredicts $\eta$. This suggests that the nonlinear spectral energy transfer is small compared to the spectral dissipation, as shown by Fig.~\ref{fig: SpectralFlux_Standing}$(b)$.}}

As discussed in previous sections, the decay phenomenology of AWT is different from that of TW and SW. Typical acoustic field $\uFluct(x,t), \pFluct(x,t)$ for a randomly initialized perturbation at a time after shock formation is shown in Fig.~\ref{fig: LengthScalesSummary}$(a)$. The velocity field corresponds to randomly positioned shocks connected with almost straight slant lines (expansion waves) and the pressure field with identical distribution of shocks but connected with horizontal lines. Shocks traveling in the same direction collide inelastically and coalesce, while those traveling in opposite directions pass through. As discussed previously, the integral length scale $\ell$ defines the average distance between the adjacent shock traveling in the same direction. Due to gradual coalescence of the shocks, $\ell$ increases in time. Moreover, as $t\rightarrow \infty$, it is obvious that two opposite traveling shocks remain in the domain and $\ell \rightarrow L$. We note that such behaviour is similar to the Burgers turbulence~\citep{Kida_1979_JFM}. Figure~\ref{fig: SpectralEnergyTurbulence} shows the fully developed compensated spectra at scaled dimensionless time $t/\tau = t_0\approx 6\times10^{-4}$. For AWT, the compensated energy spectrum $\widehat{E}_kk^{2}\epsilon^{-2/3}\ell^{1/3}$ defined in Eq.~\eqref{eq: SpectralForm} does not remain constant in the energy transfer range of wavenumbers due to decay laws of energy and dissipation derived in the previous section.{\color{black}{ Moreover, for lowest $\mathrm{Re}_L$ case, the spectra exhibit energy for $k\eta>1$ (Fig.~\ref{fig: SpectralEnergyTurbulence}$(a)$) due to underprediction of $\eta$ obtained via balancing of nonlinear wave propagation and thermoviscous dissipation effects. The dissipation acts at large length scales also in the lowest $\mathrm{Re}_L$ case. Consequently, the spectral energy flux $\widehat{\Pi}_k$ is very small compared to dissipation $\epsilon$ and the length scale $\eta$ is primarily governed by diffusion only.}}

\begin{figure}
\centering
\includegraphics[width=\linewidth]{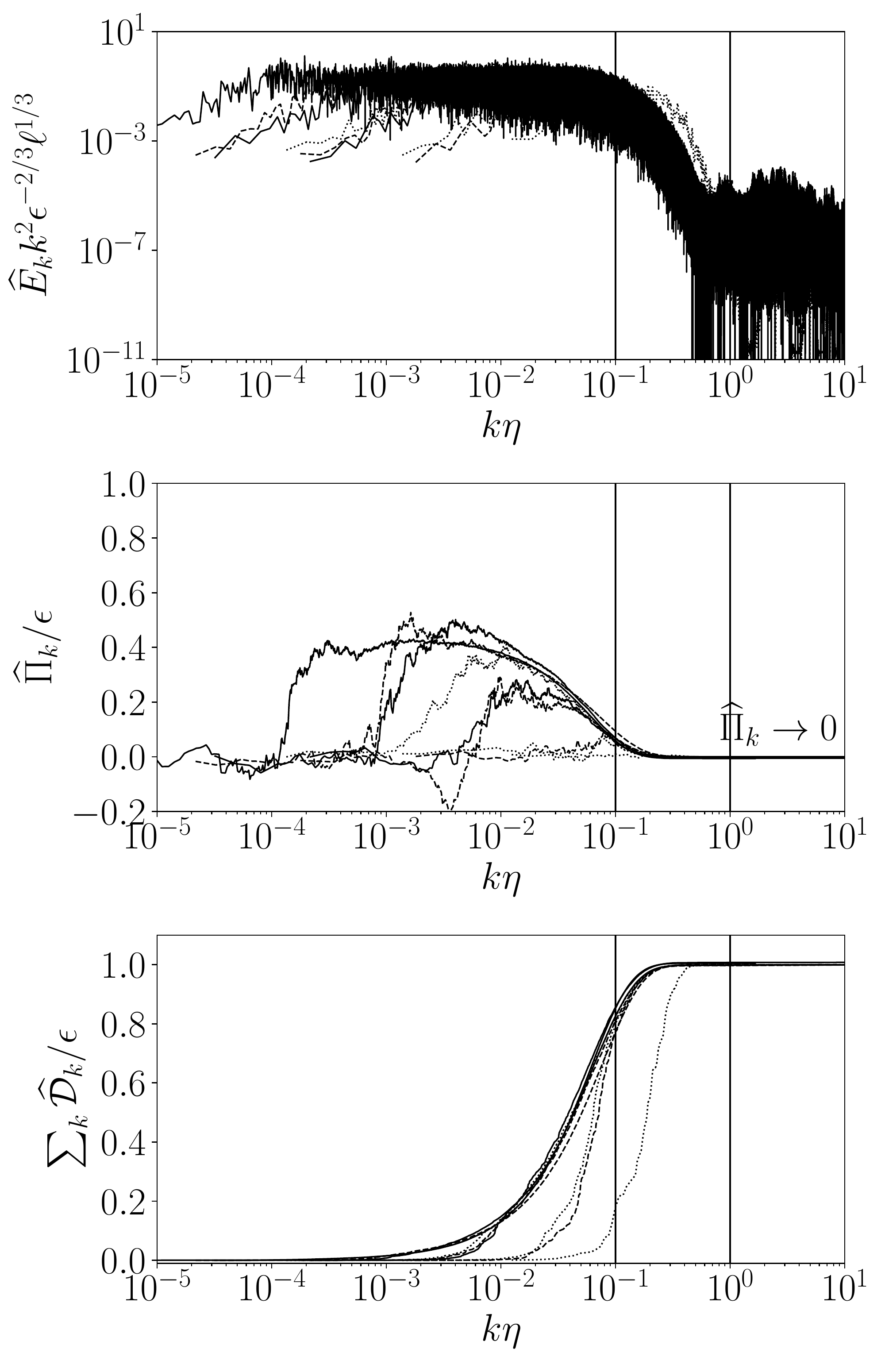}
\put(-240,370){$(a)$}
\put(-240,250){$(b)$}
\put(-240,120){$(c)$}
\caption{Fully developed spectra of compensated energy, $(a)$, spectral energy flux $(b)$, and cumulative dissipation $(c)$ against scaled wavenumber $k\eta$ for the randomly initialized broadband noise (AWT) cases with $A_{\mathrm{rms},0}$ and $\nu_0$ listed in Table~\ref{tab: test_cases} at dimensionless time $t/\tau = \tau_0\approx 6\times10^{-4}$. The marked regime $0.1<k\eta<1$ signifies the dissipation range. (--) $A_{\mathrm{rms},0}=10^{-1}$; ($--$) $A_{\mathrm{rms},0}=10^{-2}$; ($\cdots$) $A_{\mathrm{rms},0}=10^{-3}$}
\label{fig: SpectralEnergyTurbulence}
\end{figure}\section{Concluding remarks}
We have studied the spectral energy transport and decay of finite amplitude planar nonlinear acoustic perturbations governed by fully compressible 1D Navier-Stokes equations through shock-resolved direct numerical simulations (DNS) focusing on propagating single harmonic traveling wave (TW), standing wave (SW), and randomly initialized Acoustic Wave Turbulence (AWT). The maximum entropy perturbations scale as $\pFluct^2$ for normalized pressure perturbation $\pFluct \sim \mathcal{O}\left(10^{-3} - 10^{-1}\right)$. Consequently, the second order nonlinear acoustic equations are adequate to derive physical conclusions on spectral energy transfer in the system. Utilizing the second order equations, we derived the analytical expression for corrected energy corollary for finite amplitude acoustic perturbations yielding infinite order correction term in the perturbation energy density. We have shown that the spatial average of the corrected perturbation energy density can be classified as a Lyapunov function for the second order nonlinear acoustic system with strictly monotonic behaviour in time. 

Utilizing the corrected energy corollary, we derived the expressions for spectral energy, spectral energy flux, and spectral dissipation, analogous to the spectral energy equation studied in hydrodynamic turbulence. Utilizing the spectral expressions, we performed theoretical study of three possible length scales characterizing a general nonlinear acoustic system, namely, the integral length scale $\ell$, the Taylor microscale $\lambda$, and the Kolmogorov length scale $\eta$. 

In traveling waves (TW) and standing waves (SW), $\ell$ remains constant in the decaying regime. Spatial average of perturbation energy decays as $\left\langle E^{(2)}\right\rangle\sim t^{-2}$ and dissipation rate as $\epsilon \sim t^{-3}$ in time. The Kolmogorov scale increases linearly in time ($\eta\sim t$) in the decaying regime. Moreover, the spectral energy for both traveling and standing waves assumes the self-similar form: $\widehat{E}_k k^2 \epsilon^{-2/3}\ell^{1/3}\sim 0.075 f(k\eta)$. 

In acoustic wave turbulence (AWT), due to gradual increase of the integral length scale $\ell$ caused by the shock coalescence, the approximate decay laws are $\left\langle E^{(2)}\right\rangle \sim t^{-2/3}$ and $\epsilon \sim t^{-5/3}$, similar to the Burgers turbulence~\citep{Burgers1974nonlinear}. While, various cases for AWT qualitatively collapse with the scaling $\widehat{E}_k k^2 \epsilon^{-2/3}\ell^{1/3}$, quantitative scaling can only be obtained utilizing a statistically stationary ensemble of shock waves combined with random forcing, which falls beyond the current scope.

\section{Acknowledgement}
We acknowledge the financial support received from the NSF/DOE under Grant No. DE-SC0018156 and Lynn Fellowship at Purdue University. Computations have been run on the high-performance computing resources provide by the Rosen Center for Advanced Computing (RCAC) at Purdue University.

\appendix
\section{Derivation of Second-Order Acoustics Equations; Role of the Thermal Equation of State}
\label{sec: appedixA}
{For a chemically inert generic gas, infinitesimal changes in dimensionless density $\rho (p,s)$ in terms of pressure $p$ and entropy $s$ are given by,
\begin{align}
d\rho &= \left(\frac{\partial \rho}{\partial p}\right)_{s} dp + \left(\frac{\partial \rho}{\partial s}\right)_{p} ds,\nonumber\\
& = \frac{\rho}{\gamma p } dp -\left(\frac{\rho^{*}_0T^*_0R^*}{\gamma p^*_0}\right)\frac{\rho^2 T}{p}\left(\frac{\gamma - 1}{\gamma}\right)ds.
\label{eq: ConstitutiveEquationAPP}
\end{align}
Substituting the above relation in the dimensionless continuity Eq.~\eqref{eq: PressureStep1}, we obtain, 
\begin{align}
\frac{\partial p}{\partial t} &+ u\frac{\partial p}{\partial x} + \gamma p\frac{\partial u}{\partial x} \nonumber \\
&= \left(\frac{\rho^{*}_0T^*_0R^*}{p^*_0}\right)\left(\frac{\gamma - 1}{\gamma}\right){\rho T}\left(\frac{\partial s}{\partial t} + u\frac{\partial s}{\partial x}\right).
\label{eq: PressureStepAPP1}
\end{align}
Non-dimensionalizing the entropy Eq.~\eqref{eq: NS_eqns3} utilizing the Eq.~\eqref{eq: per_norm}, we obtain, 
\begin{equation}
\rho T \left(\frac{\partial s}{\partial t} + u\frac{\partial s}{\partial x}\right) = \frac{\nu_0}{Pr}\frac{C^*_p}{R^*}\frac{\partial^2 T}{\partial x^2} + \frac{4\nu_0}{3}\frac{a^{*2}_0}{R^*T^*_0}\left(\frac{\partial u}{\partial x}\right)^2.
\label{eq: NonDimEnt}
\end{equation}
Substituting the above equation in Eq.~\eqref{eq: PressureStepAPP1}, we obtain, 
\begin{align}
&\frac{\partial p}{\partial t} + u\frac{\partial p}{\partial x} + \gamma p\frac{\partial u}{\partial x} \nonumber \\
&= \left(\frac{\rho^{*}_0T^*_0R^*}{ p^*_0}\right)\frac{\gamma - 1}{\gamma }\left(\frac{\nu_0}{Pr}\frac{C^*_p}{R^*}\frac{\partial^2 T}{\partial x^2} + \frac{4\nu_0}{3}\frac{a^{*2}_0}{R^*T^*_0}\left(\frac{\partial u}{\partial x}\right)^2\right).
\label{eq: PressureStepAPP2}
\end{align}
Substituting the decomposition of variables (cf. Eq.~\eqref{eq: per_norm}) in the above Eq.~\eqref{eq: PressureStepAPP2}, we obtain the pressure perturbation equation for a generic gas, 
\begin{align}
&\frac{\partial \pFluct}{\partial t} + \frac{\partial \pFluct}{\partial x} + \uFluct\frac{\partial \pFluct}{\partial x} + \gamma \pFluct\frac{\partial \uFluct}{\partial x} \nonumber \\
&= \left(\frac{\rho^{*}_0T^*_0R^*}{ p^*_0}\right)\frac{\gamma - 1}{\gamma }\left(\frac{\nu_0}{Pr}\frac{C^*_p}{R^*}\frac{\partial^2 T'}{\partial x^2} + \frac{4\nu_0}{3}\frac{a^{*2}_0}{R^*T^*_0}\left(\frac{\partial \uFluct}{\partial x}\right)^2\right).
\label{eq: PressureStepAPP3}
\end{align}
As shown in Section~\ref{sec: 2ndOrderEntropy}, the entropy perturbations are atmost $2^{\mathrm{nd}}$ order in pressure, independent of viscosity. Consequently, the first and second term on right hand side of Eq.~\eqref{eq: NonDimEnt} are second and third order in pressure perturbations, respectively. Truncating the Eq.~\eqref{eq: PressureStepAPP3} up to second order, we obtain the second order equation for pressure perturbations for a generic fluid as, 
\begin{align}
&\frac{\partial \pFluct}{\partial t} + \uFluct\frac{\partial \pFluct}{\partial x} + \frac{\partial \uFluct}{\partial x} + \gamma \pFluct\frac{\partial \uFluct}{\partial x} = \nonumber \\
&\frac{\nu_0}{Pr}\left(\frac{\rho^{*}_0T^*_0R^*}{ p^*_0}\right)\left(\frac{\gamma - 1}{\gamma }\right)\left(\frac{\partial T}{\partial p}\right)_{s,0}\frac{C^*_p}{R^*}\frac{\partial^2 \pFluct}{\partial x^2} \nonumber \\
&+ \mathcal{O}\left(\pFluct\sFluct, \sFluct^2, \pFluct^3, \left(\frac{\partial \uFluct}{\partial x}\right)^2\right),
\label{eq: PressureStepAPP4}
\end{align}
Substituting the decomposition of variables (cf. Eq.~\eqref{eq: per_norm}) in dimensionless Eq.~\eqref{eq: NS_eqns2} and neglecting changes in kinematic viscosity, we obtain, 
\begin{equation}
 \frac{\partial \uFluct}{\partial t} + \uFluct\frac{\partial \uFluct}{\partial x} + \frac{1}{1 + \rho'}\frac{\partial \pFluct}{\partial x} = \frac{4}{3}\nu_0\frac{\partial^2 \uFluct}{\partial x^2}.
 \label{eq: velocityStep1}
\end{equation}
Equations~\eqref{eq: PressureStepAPP4} and~\eqref{eq: velocityStep1} do not involve any assumption regarding the thermal equation of state of the gas and hold for any chemically inert generic gas.\\

Assuming a thermal equation of state for an ideal gas in Eq.~\eqref{eq: PressureStepAPP4} and utilizing binomial expansion in Eq.~\eqref{eq: velocityStep1}, we obtain Eqs.~\eqref{eq: pressure} and~\eqref{eq: velocity} as, 
\begin{align}
\frac{\partial \pFluct}{\partial t} + \frac{\partial \uFluct}{\partial x} + \gamma \pFluct \frac{\partial \uFluct}{\partial x} &+ \uFluct\frac{\partial \pFluct}{\partial x} = \nu_0\left(\frac{\gamma - 1}{Pr}\right)\frac{\partial^2 \pFluct}{\partial x^2} \nonumber \\
& + \mathcal{O}\left(\pFluct\sFluct, \sFluct^2, \pFluct^3, \left(\frac{\partial \uFluct}{\partial x}\right)^2\right),\label{eq: pressureAPP}
\end{align}
\begin{align}
\frac{\partial \uFluct}{\partial t} + \frac{\partial \pFluct}{\partial x} + \frac{\partial}{\partial x}\left(\frac{\uFluct^2}{2} - \frac{\pFluct^2}{2}\right) &= \frac{4}{3}\nu_0\frac{\partial^2 \uFluct}{\partial x^2} \nonumber \\
&+\mathcal{O}\left(\rho'^2\pFluct,\rho'^3\pFluct \right). \label{eq: velocityAPP}
\end{align}
We note that the LHS of Eqs~\eqref{eq: PressureStepAPP4} and \eqref{eq: velocityStep1} (up to second order) are identical to those of Eqs.~\eqref{eq: pressureAPP} and \eqref{eq: velocityAPP}, respectively, hence independent from the thermal equation of state. As shown in section~\ref{sec: EnergyCorollary}, the functional form of the second order perturbation energy norm $E^{(2)}$ (Eq \eqref{eq: Energy_norm}) is exclusively dictated by such terms, and hence is also independent from the thermal equation of state. The results shown in this work focus on ideal-gas simulations merely for the sake of simplicity, with no loss of generality pertaining to inviscid nonlinear (up to second order) spectral energy transfer dynamics.}

\section{Derivation of Spectral Energy Transfer}
\label{sec: appedixB}
Equation~\eqref{eq: spectral_conservation} can be obtained from the conservation of perturbation energy upon considering the second order governing relations (Eqs.~\eqref{eq: pressure} and~\eqref{eq: velocity}) and substituting the Fourier expansions of $\pFluct$ and $\uFluct$,
\begin{equation}
\pFluct = \sum^{\infty}_{k=-\infty}\widehat{p}_k e^{2\pi ikx}, ~~ \uFluct=\sum^{\infty}_{k=-\infty}\widehat{u}_k e^{2\pi ikx},
\end{equation}
yielding,
\begin{align}
\frac{d\widehat{p}_k}{dt} + &2\pi ik\widehat{u}_k + 2\pi i\gamma\sum^{\infty}_{k'=-\infty}k'\widehat{p}_{k-k'}\widehat{u}_{k'} + \dots \nonumber \\
 &2\pi i\sum^{\infty}_{k'=-\infty}k'\widehat{p}_{k'}\widehat{u}_{k-k'} = -4\pi^2\nu_0\left(\frac{\gamma - 1}{Pr}\right) k^2 \widehat{p}_k,\label{eq: p_spectral}\\
\frac{d\widehat{u}_k}{dt} + &2\pi ik\widehat{p}_k + 2\pi i\sum^{\infty}_{k'=-\infty}k'\widehat{u}_{k-k'}\widehat{u}_{k'} + \dots \nonumber \\
&- 2\pi i\sum^{\infty}_{k'=-\infty}k'\widehat{p}_{k-k'}\widehat{p}_{k'} = -\frac{16\pi^2}{3}\nu_0k^2\widehat{u}_k.\label{eq: u_spectral}
\end{align} 
Multiplying eqs.~\eqref{eq: p_spectral} and~\eqref{eq: u_spectral} by $\widehat{p}_{-k}$ and $\widehat{u}_{-k}$ and adding the complex conjugate, we obtain,
\begin{align}
&\frac{d}{dt}\left(\frac{|\widehat{p}_k|^2}{2} + \frac{|\widehat{u}_k|^2}{2}\right) + 2\pi \gamma \Re\left(\widehat{p}_{-k}\sum^{\infty}_{k'=-\infty}ik'\widehat{u}_{k'}\widehat{p}_{k-k'}\right) \nonumber \\
&+ \Re\left(\widehat{p}_{-k}\widehat{\left(u\frac{\partial p}{\partial x}\right)}_k + \frac{\widehat{u}_{-k}}{2}\reallywidehat{\left(\frac{\partial }{\partial x}\left(u^2-p^2\right)\right)}_k \right) \nonumber \\
& = -4\pi^2\nu_0\frac{\gamma - 1}{Pr} k^2|\widehat{p}_k|^2 - \frac{16\pi^2}{3}\nu_0k^2|\widehat{u}_k|^2.
\label{eq: LinEq1}
\end{align}
The second term in the above equation can be evaluated recursively utilizing the Eq.~\eqref{eq: pressure} yielding,
\begin{align}
&2\pi\gamma \Re\left(\widehat{p}_{-k}\sum^{\infty}_{k'=-\infty}ik'\widehat{u}_{k'}\widehat{p}_{k-k'}\right) = \Re\left(\widehat{p}_{-k}\frac{d\widehat{g}_k}{dt}\right) +\nonumber\\
&\Re\left(\widehat{p}_{-k}\widehat{\left(\frac{\partial ug}{\partial x}\right)}_k\right) - \nu_0\left(\frac{\gamma - 1}{Pr}\right)\Re\left(\widehat{p}_{-k}\widehat{\left(\frac{\partial g}{\partial p}\frac{\partial^2 p}{\partial x^2}\right)}_k\right),
\end{align}
which, upon substitution in Eq.~\eqref{eq: LinEq1} yields, 
\begin{equation}
\frac{d \widehat{E}_k}{dt} + \widehat{T}_k = \widehat{\mathcal{D}}_k,
\end{equation}
where, the spectral energy transfer function $T_k$ is given by, 
\begin{align}
\widehat{T}_k = \Re\Bigg(\widehat{p}_{-k}\widehat{\left(\frac{\partial ug}{\partial x}\right)}_k + \widehat{p}_{-k}\widehat{\left(u\frac{\partial p}{\partial x}\right)}_k\nonumber \\
+\frac{\widehat{u_{-k}}}{2}\reallywidehat{\left(\frac{\partial }{\partial x}\left(u^2-p^2\right)\right)_k\Bigg)}, 
\end{align}
and the spectral dissipation term $\mathcal{D}_k$ is given by,
\begin{align}
\mathcal{D}_k &= -\nu_0\left(\frac{\gamma - 1}{Pr}\right) \left(4\pi^2k^2|\widehat{p}_k|^2 - \Re\left(p_{-k}\reallywidehat{\left(\frac{\partial g}{\partial p}\frac{\partial^2 p}{\partial x^2}\right)_k} \right)\right)-\nonumber \\
 &\frac{16\pi^2}{3}\nu_0k^2|\widehat{u}_k|^2.
\end{align}
Summation of Eq.~\eqref{eq: LinEq1} for $k'<k$ yields Eq.~\eqref{eq: spectral_conservation} and the expressions thereafter.

\bibliography{references}

\end{document}